\documentclass[a4paper,fleqn,usenatbib]{mnras}

\usepackage{newtxtext,newtxmath}
\usepackage[T1]{fontenc}
\usepackage{ae,aecompl}
\usepackage{listings}

\usepackage{graphicx}	% Including figure files
\usepackage{amsmath}	% Advanced maths commands
\usepackage{amssymb}	% Extra maths symbols
\usepackage{color}      % use if color is used in text

\graphicspath{{./Graphs/}}
\def\Msol{\mathrel{\rm M_{\odot}}} % Solar mass
\def\Msolyr{\mathrel{\rm M_{\odot}yr^{-1}}} % SFR in Solar mass per year
\def\ergss{\mathrel{\rm erg \; s^{-1}}} % Ergs per s
\def\lx{L_{\rm X}} % Hard x ray lum
\def\lir{\mathrel{\rm L_{IR,SF}}} % Infra red lum due to SF
\def\lirsed{\mathrel{\rm L_{IR,SF,Sol}}} % Infra red lum due to SF for SED
\def\mSFR{\mathrel{\langle \rm SFR \rangle}}
\def\msSFR{\mathrel{\langle \rm sSFR \rangle}}

\definecolor{purple}{RGB}{175,0,175}
\definecolor{red}{RGB}{255,0,0}
\definecolor{darkblue}{RGB}{0,0,175}
\definecolor{lime}{RGB}{0,255,0}

\title[A signature of AGN Feedback]{Identifying the subtle signatures
  of feedback from distant AGN using ALMA observations and
  the EAGLE hydrodynamical simulations}

\author[J. Scholtz, et al.]{\parbox[h]{\textwidth}{ 
J.\ Scholtz$,^{\! 1}$\thanks{E-mail: honzascholtz@gmail.com}
D.\ M.\ Alexander,$^{\! 1}$
C.\ M.\ Harrison,$^{\! 2,1}$
D.\ J. Rosario,$^{\! 1}$
S.\ McAlpine,$^{\! 3}$
J.R Mullaney,$^{\! 4}$
F.\ Stanley,$^{\! 5,1}$
J.\ Simpson,$^{\! 6}$
T.\ Theuns,$^{\! 3}$
R.\ G. Bower,$^{\! 3}$
R.\ C. Hickox,$^{\! 7}$
P. \ Santini,$^{\! 8}$
and A.\ M. Swinbank.$^{\! 1}$
}
\\
\\
$^{1}$ Centre for Extragalactic Astronomy, Durham University, Department of Physics, South Road, Durham, DH1 3LE, UK\\
$^{2}$ European Southern Observatory, Karl-Schwarzschild-Str. 2, 85748 Garching b. Munchen, Germany\\
$^{3}$ Institute for Computational Cosmology, Department of Physics, Durham University, South Road, Durham, DH1 3LE, UK\\
$^{4}$ Department of Physics \& Astronomy, University of Sheffield, Hounsfield Road, Sheffield, S3 7RH, UK\\
$^{5}$ Department of Space Earth and Environment, Chalmers University of Technology, Onsala Space Observatory, SE-43992 Onsala, Sweden \\
$^{6}$ Academia Sinica Institute of Astronomy and Astrophysics (ASIAA), No. 1, Section 4, Roosevelt Rd., Taipei 10617, Taiwan\\
$^{7}$ Department of Physics and Astronomy, Dartmouth College, 6127 Wilder Laboratory, Hanover, NH 03755, USA \\
$^{8}$ INAF - Osservatorio Astronomico di Roma, via di Frascati 33, 00078 Monte Porzio Catone, Italy \\
}

\date{XYZ}

\pubyear{2017}

\begin{document}
\label{firstpage}
\pagerange{\pageref{firstpage}--\pageref{lastpage}}
\maketitle

\begin{abstract}
We present sensitive 870~$\mu$m continuum measurements from our ALMA programmes of 114 X-ray selected AGN in the CDF-S and COSMOS fields. We use these observations in combination with data from \textit{Spitzer} and  \textit{Herschel} to construct a sample of 86 X-ray selected AGN, 63 with ALMA constraints at $z=1.5-3.2$ with stellar mass $>2\times10^{10}$~$M_{\odot}$. We constructed broad-band spectral energy distributions in the infrared band (8 -- 1000~$\mu$m) and constrain star-formation rates (SFRs) uncontaminated by the AGN. Using a hierarchical Bayesian method that takes into account the information from upper limits, we fit SFR and specific SFR (sSFR) distributions. We explore these distributions as a function of both X-ray luminosity and stellar mass. We compare our measurements to two versions of the EAGLE hydrodynamical simulations: the reference model with AGN feedback and the model without AGN. We find good agreement between the observations and that predicted by the EAGLE reference model for the modes and widths of the sSFR distributions as a function of both X-ray luminosity and stellar mass; however, we found that the EAGLE model without AGN feedback predicts a significantly narrower width when compared to the data. Overall, from the combination of the observations with the model predictions, we conclude that (1) even with AGN feedback, we expect no strong relationship between the sSFR distribution parameters and instantaneous AGN luminosity and (2) a signature of AGN feedback is a broad distribution of sSFRs for all galaxies (not just those hosting an AGN) with stellar masses above $\approx 10^{10}$M$_{\odot}$.

\end{abstract}


%%%%%%%%%%%%%%%%%%%%%%%%%%%%%%%%%%%%%%%%%%%%%%%%%%

%%%%%%%%%%%%%%%%% BODY OF PAPER %%%%%%%%%%%%%%%%%%

\section{Introduction}\label{Intro}

The most successful models of galaxy formation require AGN activity
(via ``AGN feedback'') to explain many of the puzzling properties of
local massive galaxies and the intergalactic medium (IGM); e.g. the
colour bi-modality of local galaxies, the steep luminosity functions, the black hole--spheroid
relationships and the metal enrichment of the intergalactic medium
\citep[see][for reviews]{Alexander12, Fabian12, Harrison17}. The key
attribute of the AGN in these models is the injection of significant
energy into the interstellar medium (ISM), which inhibits or
suppresses star formation by either heating the ISM or ejecting the
gas out of the host galaxy through outflows
\citep{Sturm11,Fabian12,Cicone14}. In recent years it has been shown
that low-redshift ($z<1$), low-accretion rate AGN are responsible for
regulating the inflow of cool gas in massive galaxy clusters through
heating \citep[see][for review]{McNamara12}. However, despite
spectroscopic observations that have shown that energetic outflows are
a common property of luminous AGN \citep[e.g.][]{Veilleux05,Ganguly08,
  Mullaney13,Cicone14,Harrison14,Balmaverde15, Harrison16, Leung17},
we lack direct observational support that they dramatically impact on
star formation in the distant Universe ($z>$~1.5), which is a
fundamental requirement for the majority of galaxy formation
models \citep[e.g.][]{Springel05a, Vogelsberger14, Schaye15}.

With high sensitivity at infrared (IR) wavelengths, {\it Herschel} has
provided new insight into the star forming properties of distant AGN
($z>1$).\footnote{The majority of studies have used X-ray observations
  to identify AGN since they provide an efficient and near
  obscuration-independent selection \citep[see \S2 at][for an overview
    of the advantages of X-ray observations in identifying
    AGN]{Brandt15}.} The broadly accepted view is that the mean
star-formation rates (SFRs) and specific SFRs (sSFRs;
i.e.,\ SFR/stellar mass) of moderate-luminosity AGN ($L_{\rm
  X}\approx10^{43}$--$10^{44}$~erg~s$^{-1}$) are consistent with those
of the coeval star-forming galaxy population \citep[e.g. also ][]
{Lutz10, Shao10, Harrison12, Mullaney12a,Santini12, Rosario13b,
  Azadi15, Stanley15, Cowley16}. The definition of the star-forming
galaxy population in this context is that of the ``main sequence'';
i.e.,\ the redshift and stellar-mass dependent evolution of sSFRs of
star-forming galaxies \citep[e.g.,][]{Noeske07, Elbaz11, Speagle14,
  Whitaker14, Schreiber15}. To first order these results suggest a
connection between AGN activity and star formation without providing
clear evidence that moderate-luminosity AGN impact on star
formation. By contrast, mixed results we presented for luminous
AGN ($L_{\rm X}>10^{44}$~erg~s$^{-1}$), with different studies arguing
that AGN either suppress, enhance, or have no influence on star
formation when compared to moderate-luminosity AGN
\citep[e.g.][]{Harrison12,Page12,Rosario12,Rovilos12,Azadi15,Stanley15}.

The majority of the current {\it Herschel} studies suffer from at
least one of the following limitations, which hinder significant
further progress: 1) SFRs are often calculated from single-band
photometry, which doesn't account for the factor $\approx$~2--3
difference in the derived SFR between star forming galaxy templates
\citep[depending on wavelength; see ][]{Stanley16phd}, 2) a modest fraction
of X-ray AGN are detected by {\it Herschel} (often $<10$\% for X-ray
AGN at $z>1.5$), which drives the majority of studies to explore the
stacked average SFR rate, which can be strongly effected by bright
outliers (e.g.,\ see \citealt{Mullaney15} for solutions to this
problem), 3) the contribution to the IR emission from the AGN is often
not directly constrained which can be significant even for
moderate-luminosity AGN \citep[e.g.][]{Mullaney11, DelMoro13}, and 4)
upper limits on SFRs are often ignored, which will bias reported SFRs
towards high values, potentially missing key signatures of suppressed
star formation. Furthermore, since mass accretion onto black holes is
a stochastic process with a timescale shorter than that of star
formation \citep[e.g.][]{Hickox14, King15,Schawinski15, McAlpine17},
we must be cautious about what can inferred from AGN feedback using
the observed relationships between SFRs and AGN luminosities
\citep[see ][]{Harrison17}. To more completely constrain the
impact that AGN have on star formation we need to measure \textit{(s)}SFR
  distributions as a function of key properties (e.g.,\ X-ray
luminosity, stellar mass), which will provide more stringent tests of
the current models of galaxy formation and evolution \citep[e.g.][]{
  Vogelsberger14, Schaye15,Lacey16}.
  
As described above, previous studies exploring the topic of star formation in AGN typically used linear means to estimate the SFR and sSFR of the AGN population; a single parameter description of the population. However, by using ALMA data, to go deeper than is possible with Herschel data alone, we already have shown in our pilot study \citep{Mullaney15} that the linear mean is consistently higher than the mode (the most common value). A linear mean of two samples can be consistent, while their distributions can be inconsistent. In that study we showed that X-ray AGN have consistent mean sSFRs but in-consistent distributions compared to main sequence galaxies. Therefore in order to adequately describe the unique star-forming properties of a population, we must constrain the parameters (the mode and the width) of the distributions of SFR or sSFR. These values are much more powerful, than a simple linear mean, to compare between different samples and to rigorously test model predictions, see \S \ref{sec: NoAGN}.

The aim of this paper is to use sensitive ALMA observations of X-ray
AGN at $z>1.5$, in conjunction with {\it Spitzer}--{\it Herschel}
photometry, to address the challenges outlined above and answer the
question: what impact do luminous AGN have on star formation? The
significantly improved sensitivity and spatial resolution that ALMA
provides over {\it Herschel} allows for the detection of star forming
emission from galaxies at $z>1.5$ up to an order of magnitude below
the equivalent sensitivity of {\it Herschel} (see \citealt{Mullaney15}; Stanley et al, submitted). In this paper we expand on the
\cite{Mullaney15} study with additional ALMA observations of X-ray AGN
to increase the overall source statistics, particularly at the high
luminosity end (i.e.,\ $L_{\rm X}>10^{44}$~erg~s$^{-1}$). We also make
a quantitative comparison of our results to those from a leading set
of hydrodynamical cosmological simulations \citep[EAGLE; Evolution and Assembly of
GaLaxies and their Environments;][]{Schaye15}.

In \S2 we describe the data and the basic analyses used in our study,
in \S3 we present our main results, including a comparison to EAGLE,
in \S4 we discuss our results within the broader context of the impact
of AGN on the star forming properties of galaxies, and in \S5 we draw
our conclusions. We also provide in the appendix the ALMA $870 \mu$m
photometry for all of the 114 X-ray sources that were either targetted in
our ALMA programmes or serendipitously lay within the ALMA field of
view. In all of our analyses we adopt the cosmological parameters of
$\rm H_0 = 71 \ km \, s^{-1}$, $\rm \Omega_M = 0.27$, $\rm
\Omega_\Lambda = 0.73$ and assume a \citet{Chabrier03} initial mass
function (IMF).

\section{Data and basic analyses} \label{sec:Data_analysis}

In this section we describe the main sample of X-ray AGN used in our
analyses, along with the calculation of the key properties (stellar
masses, SFR and sSFR) and associated errors (see \S\ref{sec:sample}),
our approach in measuring the properties of the (s)SFR distributions
(see \S\ref{sec:Fits}), and the EAGLE hydrodynamical cosmological
simulations used to help interpret our results (see
\S\ref{sec:EAGLE_d}).

\subsection{Main sample: definition and properties}\label{sec:sample}

The prime objective of our study is to constrain the star forming
properties of X-ray AGN to search for the signature of AGN
feedback. To achieve this we 1) need to select AGN over the redshift
and luminosity ranges where AGN feedback is thought to be important
and 2) require sensitive star formation and stellar-mass
measurements. On the basis of the first requirement our main sample is
defined with the following criteria:

\begin{enumerate}
\item rest-frame 2--10 keV luminosity of $\lx = 10^{43} -10^{45} \ergss$,
\item redshift of $z=1.5-3.2$, and
\item stellar mass of $M_{*} > 2 \times 10^{10} \Msol$.
\end{enumerate}

\noindent The redshift and X-ray luminosity ranges ensure that we
include AGN that 1) are most likely to drive energetic outflows \citep{Harrison16}, and
consequently have direct impact on the star formation in the host
galaxies and 2) contribute to the majority of the cosmic black-hole
and galaxy growth \citep{Madau14,Brandt15}. The stellar-mass cut is
required since probing the star forming properties below the main
sequence for individual systems with $M_{*}< 2 \times 10^{10} \Msol$
requires deeper IR data than is currently available. Furthermore, the
cosmological simulations predict that the impact of AGN feedback is
most significant in more massive galaxies \citep[e.g.][]{Bower17,
  McAlpine17}.

Given these criteria, we selected X-ray AGN from the \textit{Chandra}
Deep Field-South (CDF-S) and the central regions of Cosmic Evolution
Survey (COSMOS), which have the deepest multi-wavelength ancillary
data available in the well-observed CANDELS (Cosmic Assembly
Near-infrared Deep Extragalactic Legacy Survey) sub regions
\citep{Grogin11,Koekemoer11}. For the CDF-S field we selected X-ray
AGN at $z=$~1.5--3.2 with $\lx = 10^{43} - 10^{44} \ergss$ from the
4~Ms \textit{Chandra} catalogues of \cite{Xue11} and \cite{Hsu14}. For
the COSMOS field we primarily selected X-ray AGN with $\lx = 10^{44} -
10^{45} \ergss$ from the central $12^{\prime}.5$-radius region using
the \textit{Chandra} catalogues of \cite{Civano16} and
\cite{Marchesi16}; however, to ensure a sufficient number of AGN at
$z=$~1.5--3.2 with $\lx = (0.3 - 1)\times10^{45} \ergss$ we expanded
the selection of the most luminous AGN to the central
$25^{\prime}$-radius region of COSMOS. Stellar mass and star formation
measurements (augmented by our sensitive ALMA observations; see
appendix) were obtained for all of the X-ray AGN that met these
criteria and the systems with $M_{*} < 2 \times 10^{10} \Msol$ were
removed; see \S\ref{Sec:Stellar_mass} and \S\ref{sec:SED} for details
of the stellar-mass and star-formation measurement procedures.

\begin{figure}
	\includegraphics[width=\columnwidth]{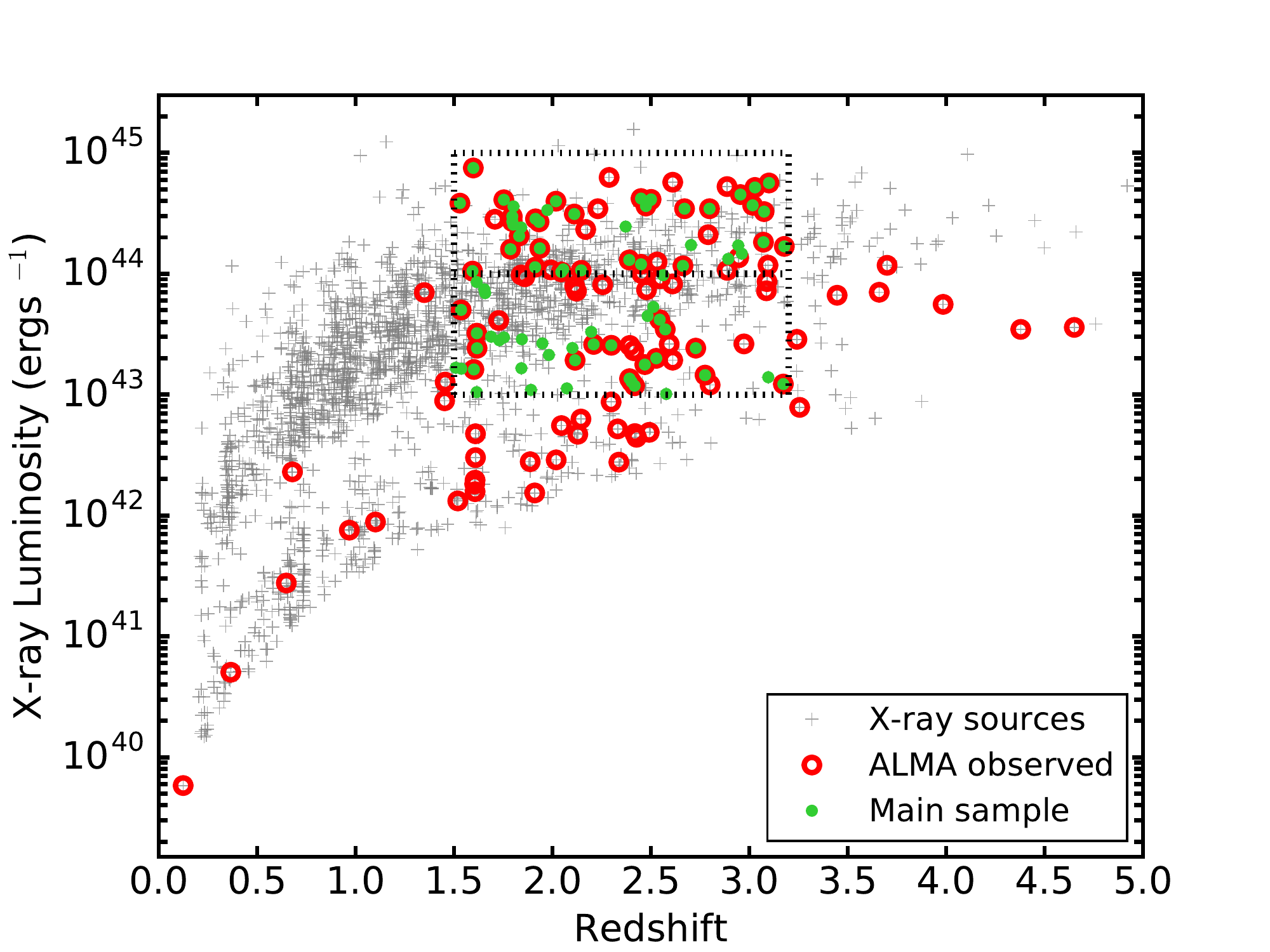}
   \caption{X-ray luminosity (2-10 keV: rest frame) versus redshift
     for the X-ray sources in the CDF-S and COSMOS fields. The X-ray
     sources that lie within our ALMA observations are indicated as
     red circles (see appendix). The X-ray AGN used in our star
     formation analyses, which comprise our main sample, are further
     highlighted with green filled circles (see \S2.1); the dotted
     square indicates the region of the X-ray luminosity--redshift
     plane used in our main analyses. Not all of the objects in the
     dotted square are selected for our main sample since many lie
     below our stellar mass threshold.}
   \label{fig:Xlum}
\end{figure}

Overall our main sample includes 81 X-ray AGN. In
Figure~\ref{fig:Xlum} we plot the X-ray luminosity versus redshift of
the overall X-ray source population in the CDF-S and COSMOS fields and
highlight the $z$--$L_{\rm X}$ parameter space explored by our main
sample. The properties of the individual X-ray AGN in the main sample
are presented in Tables \ref{Main_Sample_G} and
\ref{Main_Sample_C}. Of the 81 X-ray AGN, 63 ($\approx$~78\%) have SFR
measurements or upper limits augmented by ALMA observations. To search
for trends in the star forming properties of X-ray AGN as a function
of key properties, we also defined subsamples based on X-ray
luminosity and stellar mass: low $\lx$ ($10^{43} - 10^{44} \ergss$; 39
X-ray AGN), high $\lx$ ($10^{44} - 10^{45} \ergss$; 42 X-ray AGN), low
mass ($2 \times 10^{10} - 8 \times 10^{10} \Msol$; 41 X-ray AGN), and
high mass ($8 \times 10^{10} - 1 \times 10^{12} \Msol$; 40 X-ray
AGN). We note that the mean and median redshifts of the $\lx$ and
stellar mass subsamples are well matched: $\delta z=$~0.1 for the
$\lx$ subsamples and $\delta z=$~0.05 for the stellar mass subsamples.

\subsubsection{Stellar mass measurements}\label{Sec:Stellar_mass}

The stellar masses of the X-ray AGN were calculated by performing SED
fitting on the broad-band UV-MIR photometry ($0.1$--$24$ $\mu$m) from
archival catalogs in the CDF-S and COSMOS fields.  For the sources in
the CDF-S field, we used the multi-wavelength catalogue of
\citet{Guo13}, which covers the CANDELS GOODS-S Deep+Wide+ERS area. A
fraction ($\approx$~33\%) of our targets lie outside the CANDELS
footprint; for these, we included photometry from the MUSYC ECDFS
catalog of \citet{Cardamone10}. For the sources in the COSMOS field,
we used the multi-wavelength catalogue of
\citet{laigle16}. Catalogue-specific procedures were used to convert
tabulated aperture photometry to zero-point corrected total
photometry. In both fields, we used \textit{Spitzer} MIPS 24 $\mu$m
photometry from \citet{Floch09} and the PEP survey \citep{lutz11} to
extend the SEDs into the observed MIR.

We modelled the broad-band SEDs of the X-ray AGN using the CIGALE
package \citep[v0.8.1,][]{burgarella05, ciesla15}.  The SEDs were
fitted using combinations of stellar and AGN emission templates. The
population synthesis models of \citet{bruzual03} represented the
stellar emission, to which dust extinction was applied following the
power-law prescription of \citet{charlot00}. The AGN emission was
modelled on the library of \citet{fritz06}, which takes a fixed shape
power-law SED representing an accretion disc, and geometry-dependent
dust emission from a smooth AGN torus. After an examination of the
entire \citet{fritz06} library, we adopted a subset of the AGN
templates (described below) that reproduce empirical AGN IR SEDs
\citep[e.g.;][]{Mullaney11,Mor&Netzer12}. We fixed the power-law
indices that describe the radial and polar dust density distribution
in the torus to 0.0 and 6.0, implying a uniform density torus that has
a sharp gradient with elevation. We assumed a single value of 150.0
for the ratio between the outer radius and inner (sublimation) radius
of the torus, and allowed for three values of the 9.7 $\mu$m Si
optical depth (0.1, 1.0, 3.0). We allowed for the full range in torus
inclination angles with respect to the line of sight and set the
normalisation of the torus models to run through the MIPS 24 $\mu$m
photometric point.

From the posterior distributions of stellar mass for each galaxy
computed using CIGALE, we calculated the median stellar mass and the
16$^{\rm th}$ and 84$^{\rm th}$ percentile values as a measure of the
uncertainty on the stellar mass; see Tables~1 \& 2.

\subsubsection{Star-formation measurements}\label{sec:SED}

The star forming properties of the X-ray AGN were calculated from
\textit{Spitzer}-IRAC $8 \mu$m, \textit{Spitzer}-IRS $16 \mu$m,
\textit{Spitzer}-MIPS $24 \mu$m, deblended \textit{Herschel}-PACS (70,
100, 160 $\mu$m), deblended \textit{Herschel}-SPIRE (250, 350, 500
$\mu$m) and our ALMA photometry ($870 \mu$m, see appendix for more
details). The \textit{Spitzer} and \textit{Herschel} photometry were
taken from the same catalogues as for our earlier \citet{Stanley15}
study: the \textit{Spitzer} IRAC and IRS data is from
\cite{Sanders07}, \cite{Damen11} and \citet{Teplitz11} for the CDF-S,
COSMOS, and GOODS-S fields, respectively. The deblended photometry
consists of the MIPS $24 \mu$m and the PACS bands from
\citet{Magnelli13}\footnote{\citet{Magnelli13} published the PACS
  catalogues for GOODS-S. The catalogue for the COSMOS field was
  created using the same method and is available to download at
  http://www.mpe.mpg.de/ir/Research/PEP/DR1.} and SPIRE photometry
from \citet{Swinbank14}. For the objects that were undetected in the
\textit{Spitzer} and \textit{Herschel} maps, we calculated $3 \sigma$
upper limits.

We used SED decomposition techniques to separate the AGN and
star-forming components from the total IR SED. The full SED fitting
procedure is presented in Stanley et al. (submitted); however, we
provide brief details here and note that we used a slightly modified
approach to obtain the final SFR values and errors for application in our sSFR
distribution fitting (see \S \ref{sec:Fits}). The SED fitting
procedure is based on \citet{Stanley15}, which fitted AGN and star
forming templates to \textit{Spitzer} and \textit{Herschel} photometry
but is updated to include ALMA continuum measurements. The AGN and 5
of the 6 star forming templates are from \citet{Mullaney11} but
extrapolated to $3-1000 \mu$m by \citet{DelMoro13}, while a $6^{\rm
  th}$ star forming template is the Arp220 galaxy template from
\citet{Silva98}, which represents an extremely dusty star forming
galaxy. The photometric measurements, uncertainties, and upper limits
were taken into account when fitting the IR SEDs. Two sets of
best-fitting SED solutions were calculated for each X-ray AGN, giving
12 best-fitting SED solutions overall: one set using each of the 6
star forming templates and the other set using the 6 star forming
templates plus the AGN template. To determine whether the fit requires
an AGN component or not, we used the Bayesian Information Criteria
\citep[BIC;][]{Schwarz78} which allows for an objective comparison
between non-nested models with a fixed data set (see section
2.3.2). To establish if the fit of the source requires an AGN
component, the SED with the AGN component has to have a smaller BIC
than that of the SED with no AGN component with a difference of
$\Delta$BIC>2 (for more information and examples see \S 3 of Stanley
et al., submitted). This way we obtain 6 SED solutions.

We integrated each star forming template from each of the 6 SED
solutions to estimate the total IR luminosities due to star formation
for that SED solution ($\lirsed$). Using this procedure we obtained 6
different values of $\lirsed$ and their errors from the fitting
routine. The final value of the IR luminosity due to star formation
($\lir$) and its error is calculated using the Bootstrap method. To
each value of $\lirsed$ we assigned a probability P($\chi^2$) (in the
shape of the $\chi^{2}$ distribution) that it is the true value of
$\lir$. Then we picked a $\lirsed$ based on its P($\chi^2$) and drew a
value of $\lir$ from a normal distribution with the mean and width as
the best value and error returned from $\lirsed$. We repeated this
procedure $10^{5}$ times to build a distribution of all possible
values of $\lir$. The created distribution was dominated by the
template with the least $\chi^2$ value, but it also took into
consideration other template solutions. For the upper limit
calculations, we selected an SED solution with the highest value of
$\lirsed$.

We converted $\lir$ to SFR using Equation 4 from \citet{Kennicutt98}
corrected to the \citet{Chabrier03} IMF. In order to calculate the
sSFR we also created a distribution of stellar masses for each object
by drawing $10^{5}$ times from the normal distribution with the mean
and width as the best value and error returned from CIGALE (see \S
\ref{Sec:Stellar_mass}). We then calculated the sSFR by dividing draws
of SFR by the draws of stellar mass. We calculated the final (and
adopted) values of the SFR and sSFR and their errors as the median and
standard deviation of the $10^{5}$ SFR and sSFR values, respectively;
see Tables~1 \& 2.

With ALMA photometry the fraction of AGN with SFR measurement increased for the low and high $\lx$ subsamples from $7 \%$ and $17 \%$ to $31 \%$ and $38 \%$, respectively (described in detail in Stanley et al, submitted). Also for those objects which remained with a SFR upper limit even with ALMA photometry, the SFR upper limits have decreased by up to factor of $10$ (Stanley et al, submitted). This significantly increased detection fraction and improved upper limits allows us to estimate the specific star-formation distributions, which was not possible without the ALMA data (see \S \ref{sec:Fits}).

\begin{table*}
 \caption{X-ray selected AGN in the main sample from the CDF-S
   field. The columns show the X-ray ID, optical position, redshift (2
   and 3 decimal places indicate photometric and spectroscopic
   redshifts, respectively), X-ray luminosity (rest-frame 2-10 keV)
   \citep[all from][]{Hsu14}, the estimated SFR from our IR SED
   fitting (see \S\ref{sec:SED}, the estimated stellar mass from our
   UV--MIR SED fitting (see \S\ref{Sec:Stellar_mass}), and a flag to
   indicate whether the X-ray AGN was observed with ALMA (see
   Table~A1).}  
   \begin{tabular}{@{}lccccccc@{}} 
\hline 
\hline 
X-ray ID&RA&Dec&Redshift& log$_{10}$&log$_{10}$& log$_{10}$&Observed\\
 &(J2000)&(J2000)& &(L$_{2-10 \space \rm{keV}}$/erg s$^{-1}$) & (SFR/M$_{\odot}$yr$^{-1})$ & (M$_{*}$/M$_{\odot}$)&with ALMA?\\
\hline 
 88&$53.01025$&$-27.76681$&$1.616$&$43.5$&$2.30\pm0.04$&$10.99\pm0.19$&yes\\
 93&$53.01271$&$-27.74731$&$2.573$&$43.5$&$<1.81$&$10.97\pm0.21$&yes\\
 111&$53.02229$&$-27.77890$&$2.51$&$43.7$&$1.83\pm0.04$&$11.28\pm0.23$&no\\
 117&$53.02548$&$-27.82436$&$1.69$&$43.5$&$1.83\pm0.16$&$10.97\pm0.15$&no\\
 142&$53.03637$&$-27.66547$&$1.54$&$43.2$&$1.69\pm0.18$&$10.84\pm0.21$&no\\
 166&$53.04548$&$-27.73749$&$1.615$&$43.9$&$2.27\pm0.02$&$10.46\pm0.17$&no\\
 176&$53.04905$&$-27.77449$&$1.51$&$43.2$&$2.03\pm0.04$&$10.35\pm0.15$&no\\
 188&$53.05392$&$-27.87690$&$2.562$&$44.0$&$<1.81$&$10.49\pm0.21$&no\\
 199&$53.05791$&$-27.83357$&$2.42$&$43.1$&$<2.25$&$11.40\pm0.16$&yes\\
 211&$53.06195$&$-27.85111$&$1.60$&$43.2$&$1.71\pm0.17$&$10.71\pm0.15$&yes\\
 213&$53.06240$&$-27.70691$&$1.891$&$43.0$&$<2.20$&$11.79\pm0.16$&no\\
 215&$53.06331$&$-27.69971$&$2.402$&$43.1$&$<1.68$&$10.86\pm0.23$&yes\\
 222&$53.06595$&$-27.70185$&$2.07$&$43.1$&$<1.69$&$11.10\pm0.23$&no\\
 240&$53.07128$&$-27.69358$&$2.20$&$43.5$&$<2.21$&$10.81\pm0.22$&no\\
 257&$53.07645$&$-27.84873$&$1.536$&$43.7$&$<2.07$&$11.17\pm0.23$&yes\\
 277&$53.08318$&$-27.71205$&$2.21$&$43.4$&$<2.20$&$10.45\pm0.23$&yes\\
 290&$53.08738$&$-27.92962$&$2.54$&$43.6$&$<1.49$&$11.04\pm0.24$&yes\\
 301&$53.09235$&$-27.80322$&$2.47$&$43.2$&$<2.41$&$10.92\pm0.22$&yes\\
 310&$53.09408$&$-27.80419$&$2.39$&$43.1$&$<1.64$&$10.68\pm0.23$&yes\\
 344&$53.10491$&$-27.70528$&$1.617$&$43.4$&$<1.76$&$11.22\pm0.15$&yes\\
 359&$53.10816$&$-27.75405$&$2.728$&$43.4$&$1.84\pm0.07$&$10.56\pm0.18$&yes\\
 369&$53.11110$&$-27.67038$&$1.658$&$43.8$&$1.65\pm0.08$&$10.49\pm0.22$&no\\
 410&$53.12414$&$-27.89127$&$2.53$&$43.3$&$2.24\pm0.12$&$11.13\pm0.17$&yes\\
 440&$53.13244$&$-27.95390$&$2.10$&$43.4$&$<2.10$&$10.68\pm0.20$&no\\
 443&$53.13366$&$-27.69865$&$1.982$&$43.3$&$<1.85$&$10.83\pm0.20$&no\\
 450&$53.13639$&$-27.86421$&$1.95$&$43.4$&$<1.92$&$11.24\pm0.17$&no\\
 456&$53.13805$&$-27.86831$&$3.17$&$43.1$&$<1.84$&$10.68\pm0.23$&yes\\
 466&$53.14169$&$-27.81662$&$2.78$&$43.2$&$<1.87$&$10.73\pm0.19$&yes\\
 486&$53.14670$&$-27.88834$&$1.84$&$43.5$&$2.19\pm0.03$&$10.41\pm0.21$&no\\
 490&$53.14883$&$-27.82112$&$2.578$&$43.0$&$<1.77$&$11.24\pm0.24$&no\\
 522&$53.15850$&$-27.77403$&$2.12$&$43.3$&$<1.83$&$10.38\pm0.24$&yes\\
 524&$53.15959$&$-27.93142$&$3.10$&$43.1$&$2.69\pm0.04$&$11.49\pm0.21$&no\\
 549&$53.16557$&$-27.76979$&$1.754$&$43.5$&$<2.54$&$10.81\pm0.22$&no\\
 575&$53.17935$&$-27.81251$&$1.730$&$43.4$&$<2.03$&$10.75\pm0.18$&no\\
 620&$53.19608$&$-27.89264$&$2.48$&$43.7$&$<1.72$&$10.86\pm0.20$&no\\
 625&$53.19886$&$-27.84391$&$1.615$&$43.0$&$<2.20$&$11.06\pm0.18$&no\\
 633&$53.20492$&$-27.91801$&$2.30$&$43.4$&$2.15\pm0.02$&$10.59\pm0.20$&yes\\
 663&$53.22878$&$-27.75165$&$1.84$&$43.2$&$<1.85$&$11.21\pm0.24$&no\\
 683&$53.24718$&$-27.81631$&$1.65$&$43.9$&$<2.13$&$11.35\pm0.18$&no\\
\hline 
\end{tabular}
\label{Main_Sample_G}
\end{table*}

\begin{table*}
 \caption{X-ray selected AGN in our main sample from the COSMOS
   field. The columns show the X-ray ID, optical position, redshift (2
   and 3 decimal places indicate photometric and spectroscopic
   redshifts, respectively), X-ray luminosity (rest-frame 2-10 keV)
   \citep[all from][]{Marchesi16}, SFR from our IR SED fitting (see
   \S\ref{sec:SED}), stellar mass from our UV--MIR SED fitting (see
   \S\ref{Sec:Stellar_mass}), and a flag to indicate whether the X-ray
   AGN was observed with ALMA (see Table~A2).}
 \begin{tabular}{@{}lccccccc@{}} 
\hline 
\hline 
X-ray ID&RA&Dec&Redshift& log$_{10}$&log$_{10}$& log$_{10}$&Observed\\
 &(J2000)&(J2000)& &(L$_{2-10 \space \rm{keV}}$/erg s$^{-1}$) & (SFR/M$_{\odot}$yr$^{-1})$ & (M$_{*}$/M$_{\odot}$)&with ALMA?\\
\hline 
cid $434$&$149.72072$&$2.34901$&$1.530$&$44.6$ &$<1.63$&$11.70\pm0.18$&yes\\
cid $580$&$149.85469$&$2.60694$&$2.11$&$44.5$ &$<1.81$&$11.13\pm0.22$&yes\\
cid $558$&$149.88252$&$2.50513$&$3.10$&$44.8$ &$1.53\pm0.18$&$11.42\pm0.21$&yes\\
cid $330$&$149.95583$&$2.02806$&$1.753$&$44.6$ &$<1.65$&$10.72\pm0.26$&yes\\
cid $2177$&$149.96660$&$2.43247$&$2.89$&$44.1$ &$1.63\pm0.07$&$11.20\pm0.23$&no\\
cid $529$&$149.98158$&$2.31501$&$3.017$&$44.6$ &$<1.80$&$11.43\pm0.20$&yes\\
cid $474$&$149.99390$&$2.30146$&$1.796$&$44.5$ &$1.11\pm0.27$&$10.38\pm0.20$&yes\\
cid $451$&$150.00253$&$2.25863$&$2.450$&$44.6$ &$1.14\pm0.19$&$11.19\pm0.19$&yes\\
cid $1127$&$150.01057$&$2.26939$&$2.390$&$44.1$ &$<1.49$&$11.02\pm0.19$&yes\\
cid $532$&$150.01985$&$2.34914$&$1.796$&$44.4$ &$<1.82$&$11.49\pm0.23$&yes\\
cid $1216$&$150.02008$&$2.35365$&$2.663$&$44.1$ &$<1.86$&$10.69\pm0.20$&yes\\
cid $659$&$150.03290$&$2.45859$&$2.045$&$44.0$ &$1.29\pm0.12$&$10.89\pm0.19$&yes\\
cid $1214$&$150.03677$&$2.35852$&$1.59$&$44.0$ &$<1.62$&$10.97\pm0.21$&yes\\
cid $351$&$150.04262$&$2.06329$&$2.018$&$44.6$ &$<1.62$&$11.15\pm0.15$&yes\\
cid $443$&$150.04597$&$2.20114$&$2.704$&$44.2$ &$<1.81$&$10.95\pm0.18$&no\\
cid $458$&$150.05524$&$2.14317$&$1.974$&$44.5$ &$1.27\pm0.18$&$10.83\pm0.25$&no\\
cid $352$&$150.05891$&$2.01518$&$2.498$&$44.6$ &$1.41\pm0.04$&$10.83\pm0.23$&yes\\
cid $1215$&$150.06454$&$2.32905$&$2.450$&$44.1$ &$<1.46$&$11.00\pm0.24$&yes\\
cid $72$&$150.09154$&$2.39908$&$2.475$&$44.6$ &$<1.85$&$10.99\pm0.22$&yes\\
cid $466$&$150.10094$&$2.16782$&$2.055$&$44.0$ &$<1.44$&$10.75\pm0.17$&no\\
cid $149$&$150.10371$&$2.66577$&$2.955$&$44.7$ &$<1.83$&$11.06\pm0.27$&yes\\
cid $1144$&$150.10477$&$2.24364$&$1.912$&$44.1$ &$<1.64$&$10.86\pm0.24$&yes\\
cid $86$&$150.11958$&$2.29591$&$1.831$&$44.3$ &$<1.46$&$11.40\pm0.18$&yes\\
cid $87$&$150.13304$&$2.30328$&$1.598$&$44.9$ &$1.53\pm0.18$&$11.52\pm0.22$&yes\\
cid $965$&$150.15218$&$2.30785$&$3.178$&$44.2$ &$1.41\pm0.19$&$10.83\pm0.17$&yes\\
cid $914$&$150.18001$&$2.23128$&$2.146$&$44.0$ &$1.60\pm0.18$&$10.90\pm0.17$&yes\\
cid $124$&$150.20532$&$2.50293$&$3.07$&$44.3$ &$<1.80$&$10.79\pm0.16$&yes\\
cid $83$&$150.21416$&$2.47502$&$3.075$&$44.5$ &$<1.83$&$11.21\pm0.20$&yes\\
cid $21$&$150.21466$&$2.20428$&$1.841$&$44.4$ &$1.50\pm0.22$&$10.41\pm0.30$&no\\
cid $23$&$150.22403$&$2.27080$&$2.944$&$44.2$ &$1.26\pm0.24$&$11.88\pm0.19$&no\\
cid $127$&$150.22702$&$2.53761$&$1.801$&$44.4$ &$2.08\pm0.08$&$11.12\pm0.23$&no\\
cid $954$&$150.23180$&$2.36401$&$1.936$&$44.2$ &$<1.83$&$10.64\pm0.30$&yes\\
cid $970$&$150.23550$&$2.36176$&$2.501$&$44.6$ &$<2.20$&$11.30\pm0.17$&yes\\
cid $75$&$150.24779$&$2.44215$&$3.029$&$44.7$ &$2.73\pm0.05$&$10.87\pm0.20$&yes\\
cid $725$&$150.27097$&$2.36507$&$2.962$&$44.2$ &$<2.42$&$10.73\pm0.16$&no\\
cid $89$&$150.28117$&$2.41590$&$2.372$&$44.4$ &$2.69\pm0.05$&$10.69\pm0.22$&no\\
cid $90$&$150.28482$&$2.39505$&$1.932$&$44.4$ &$<2.11$&$11.29\pm0.25$&yes\\
cid $365$&$150.28563$&$2.01459$&$2.671$&$44.5$ &$<2.55$&$10.62\pm0.20$&yes\\
cid $94$&$150.30956$&$2.39915$&$1.802$&$44.6$ &$<2.26$&$11.01\pm0.18$&no\\
cid $58$&$150.32689$&$2.09415$&$2.798$&$44.5$ &$<2.41$&$11.89\pm0.23$&yes\\
cid $53$&$150.34372$&$2.14067$&$1.787$&$44.2$ &$2.48\pm0.06$&$11.09\pm0.20$&yes\\
cid $62$&$150.37364$&$2.11203$&$1.914$&$44.5$ &$<2.48$&$10.51\pm0.30$&yes\\
\hline 
\end{tabular}
\label{Main_Sample_C}
\end{table*}

\subsection{Measuring the (specific) star-formation distributions}\label{sec:Fits}

The majority of previous studies have explored the mean SFRs and sSFRs
of X-ray AGN. However, the mean is sensitive to bright outliers and
can hide subtle trends in the data. A more comprehensive approach to
characterising the star forming properties of X-ray AGN, is the
measurement of the \textit{distributions} of SFRs and sSFRs. In our
analyses here we fitted the SFR and sSFR distributions of the X-ray
AGN assuming a log-normal function:

\begin{equation}
   N(x) \propto \exp \left(-\frac{ \log_{10} \left(\frac{x}{\mu}\right)^{2}}{2w^{2}}\right),
	\label{eq:log}
\end{equation}
where $x$ is the SFR or sSFR, $\mu$ is the mode, and $w$ is the width
of the distribution. The motivation for fitting a log-normal function
is: 1) the SFR and sSFR values for main-sequence galaxies broadly
follow this distribution \citep[e.g.][]{Schreiber15}, and 2) the SFR
and sSFR distributions of the AGN in the EAGLE simulations are
consistent with a log-normal function, as we demonstrate in
\S\ref{sec:modes}. Also, our source statistics are not high enough to
fit a more complex model with more parameters. However, even if the
log-normal distribution is not absolutely correct, it allows us to
broadly characterise the typical values and range in values to search
for trends and compare to the different models (see \S \ref{sec:
  NoAGN}).

The majority ($\approx$~65~\%) of the X-ray AGN in our main sample are
undetected by both \textit{Herschel} and ALMA and therefore only have a
SFR upper limit. The SFR and sSFR distributions cannot be obtained
trivially without the appropriate consideration of these limits.
Following \citet{Mullaney15}, we use a hierarchical Bayesian method to
find the best fitting parameters to sample the probability
distribution (PD) of our parameters $\mu$ and $w$, using Gibbs
sampling and Metropolis-Hastings Markov Chain Monte Carlo (MCMC)
algorithms. There are several advantages of this method: 1) the
uncertainties and upper limits can be taken into account, and 2) the
PD produced in this way can be used to estimate errors on $\mu$ and
$w$. The fitting routine treats upper limits and detections
differently, but in a statistically consistent way. For a detection,
we assumed that the likelihood function of the errors has a log-normal
shape, while for the upper limits we assumed that the likelihood
function is in the form of a log-error function. The final values and
errors of the mode $\mu$ and width $w$ are taken to be the median
values of the PD and the $68 \%$ confidence interval, respectively. As
was done in \citet{Mullaney15}, we assume uniform, uninformative
priors on $\mu$ and $w$ which do not influence the final PDs. We quote
the final values of our fits to the sSFR distributions for the main
sample (see \S \ref{sec:modes}) in Table \ref{modes}.

\begin{figure}
	\includegraphics[width=\columnwidth]{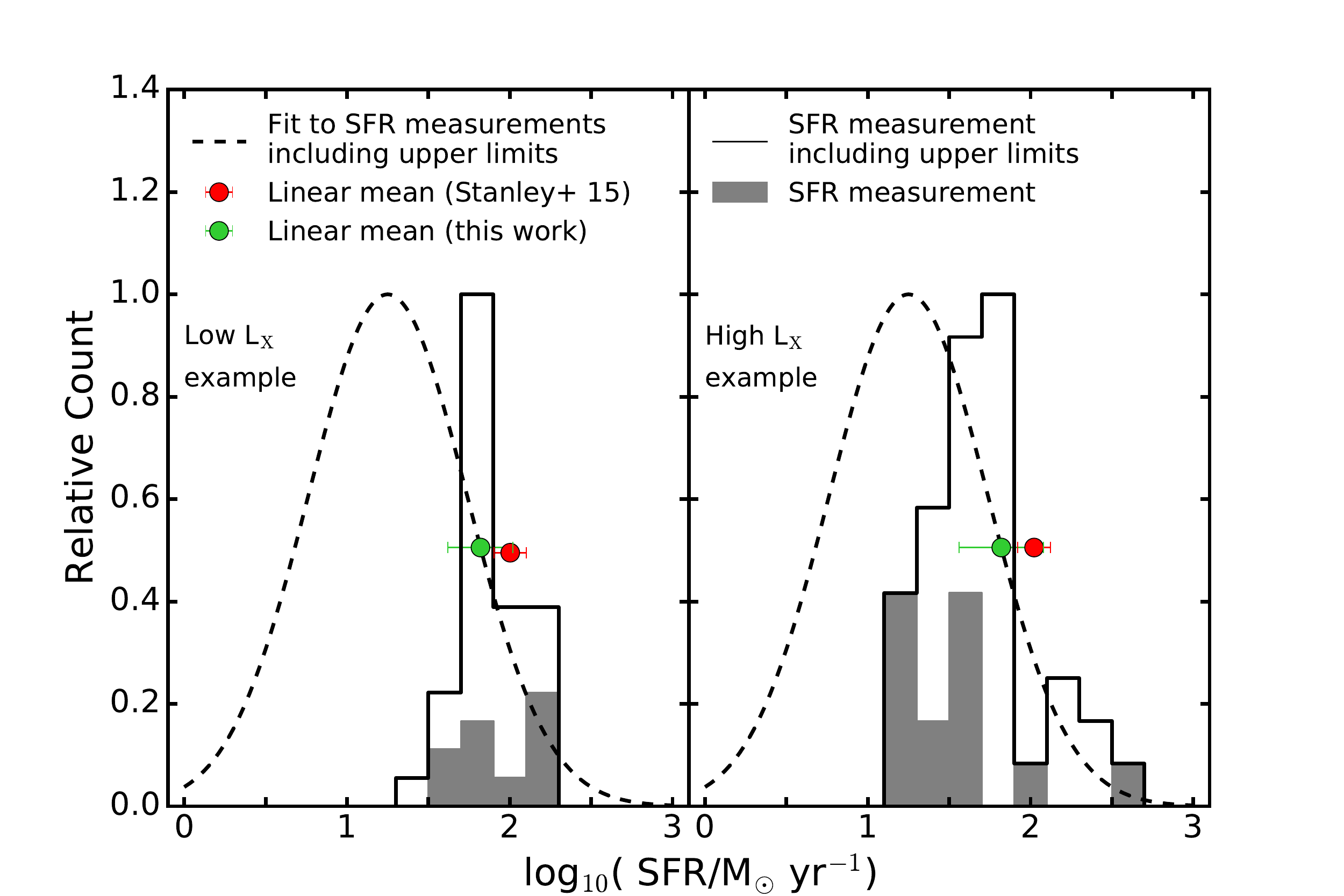}
   \caption{Example SFR distributions to demonstrate our model-fitting
     approach; see \S2.2. The X-ray AGN lie at $z=$~1.5--2.5 and have
     L$_{X} = 10^{43} - 10^{44} \ergss$ (left panel) and L$_{X} =
     10^{44} - 10^{45} \ergss$ (right panel). The filled grey
     histogram indicates the distribution of SFR measurements and the
     unfilled histogram indicates the distribution of SFR measurements
     including upper limits. The dashed curve indicates the
     best-fitting log-normal distribution to the measured SFRs
     including upper limits (see \S \ref{sec:Fits}) and the filled
     green circle indicates the mean SFR calculated from the
     best-fitting distribution. The filled red circle indicates the
     mean SFR from \citet{Stanley15} for a larger sample of X-ray AGN
     at $z= $1.5--2.5 in the same $\lx$ range but with SFR constraints
     from {\it Spitzer} and {\it Herschel} data. The error bars
     represent the $68 \%$ confidence interval for each of the
     measurements.}
   \label{Figure:Comp_st}
\end{figure}

We now test whether our method and data are consistent with earlier work, in particular \citet{Stanley15}, which used the same SED-fitting code as that adopted in this study. This earlier study relied on calculating linear means of SFR and stacking and therefore only presented linear means in bins of $\lx$, with no differentiation of the sample by stellar mass. Therefore, to replicate this study in the limited range of redshift and $\lx$ of our sample, we calculate the linear means of SFR of all AGN (including those with $\rm M_{*}< 2 \times
10^{10} \Msol$) in the $z=1.5-2.5$ redshift range. This was done directly from the corresponding log-normal distributions as follows:
\begin{equation}
\langle x \rangle= 10^{(\mu + 1.15 w^2)},
\label{eq:lm}
\end{equation}
where $\mu$ is the mode and $w$ is the width of the distribution as in
Equation \ref{eq:log}. The linear mean was calculated from the PD of
$\mu$ and $w$ from our MCMC analysis, from which the median and $68\%$
confidence interval were derived.

The $\log_{10}(\mSFR/\Msolyr)$ of our low and high $\lx$ subsamples
were $1.94^{+0.33}_{-0.20}$ and $1.8^{+0.22}_{-0.15}$, respectively,
as compared to $2.00 \pm 0.10$ and $2.02 \pm 0.10$ from
\citet{Stanley15}, see Figure \ref{Figure:Comp_st}. As such, our estimates are in good agreement with
those of \citet{Stanley15} and confirms that our new method is
consistent with previous work. In comparison, the
$\log_{10}(\mu/\Msolyr)$ of the SFR distribution for low and high
$\lx$ subsamples are $1.27^{+0.31}_{-0.22}$ and
$1.12^{+0.15}_{-0.19}$, respectively. The linear mean of the SFR is
always higher (depending on the width of the distribution) than the
mode of the distribution, making the mode of the distribution a more
reliable tracer of the typical values of the population. In summary, 
our method yields consistent result with previous studies using linear means 
and stacking procedures.

\begin{table}
 \caption{Best fitting log-normal fit parameters for the sSFR
   distributions of our main sample and sample from EAGLE simulations
   binned by X-ray luminosity and stellar mass. The quoted $\mu$ and
   $w$ and their errors are the median of the their posterior
   probability distributions (PDs) and $68 \%$ confidence intervals.
   The linear mean is calculated from $\mu$ and $w$ using equation
   \ref{eq:lm}.}
\begin{center}
\resizebox{\columnwidth}{!}{\begin{tabular}{@{}lccc@{}}
\hline
\hline
Sample& Mode ($\mu$)&Width($w$)& linear mean\\
&$\log_{10}(\mu$/Gyr$^{-1}$)&(dex)&$\log_{10}(\msSFR$/Gyr$^{-1}$)\\
\hline
Main Sample (Observed AGN): &&&\\
Low $L_{x}$ AGN&$0.03^{+0.14}_{-0.17}$&$0.52^{+0.13}_{-0.10}$& $0.34  ^{+0.18} _{-0.15}$\\
High $L_{x}$ AGN&$-0.32^{+0.15}_{-0.17}$&$0.65^{+0.15}_{-0.11}$& $ 0.17  ^{+0.26} _{-0.19}$\\
\\
Low Mass AGN&$-0.01^{+0.13}_{-0.15}$&$0.53^{+0.13}_{-0.08}$& $0.31 ^{+0.16} _{-0.14}$\\ 
High Mass AGN&$-0.48^{+0.17}_{-0.20}$&$0.67^{+0.18}_{-0.12}$& $0.05  ^{+0.29} _{-0.22}$ \\
\hline
EAGLE ref model: &&&\\
Low $L_{x}$ AGN &$-0.08^{+0.05}_{-0.04}$&$0.45^{+0.06}_{-0.06}$& $0.14  ^{+0.08} _{-0.1}$\\
High $L_{x}$ AGN &$0.14^{+0.05}_{-0.04}$&$0.45^{+0.05}_{-0.04}$& $ 0.38  ^{+0.08} _{-0.07}$\\
\\
Low Mass AGN &$0.04^{+0.02}_{-0.02}$&$0.47^{+0.02}_{-0.02}$&$0.23^{+0.03}_{-0.03}$\\
High Mass AGN &$-0.23^{+0.07}_{-0.07}$&$0.42^{+0.05}_{-0.05}$&$-0.03^{+0.09}_{-0.07}$\\
\\
Low Mass galaxy&$-0.14^{+0.02}_{-0.02}$&$0.48^{+0.02}_{-0.02}$&$0.22^{+0.02}_{-0.02}$\\
High Mass galaxy&$-0.31^{+0.02}_{-0.02}$&$0.45^{+0.02}_{-0.02}$&$-0.15^{+0.02}_{-0.02}$\\

\hline
EAGLE no AGN model: &&&\\
Low Mass galaxy &$0.13^{+0.01}_{-0.01}$&$0.23^{+0.01}_{-0.01}$& $0.20 ^{+0.02} _{-0.02}$\\ 
High Mass galaxy &$-0.10^{+0.01}_{-0.01}$&$0.28^{+0.01}_{-0.01}$& $0.0 ^{+0.02} _{-0.02}$\\ 
\hline
\end{tabular}}
\end{center}
\label{modes}
\end{table}

\subsection{EAGLE hydrodynamical simulation and source properties}\label{sec:EAGLE_d}

Cosmological simulations of galaxy formation have provided some of the
most compelling evidence that AGN feedback has a significant effect on
star formation in the galaxy population. To aid in the interpretation
of our data we have therefore compared the sSFR distributions of the
X-ray AGN in our main sample to those computed from the EAGLE
cosmological hydrodynamical simulation \citep{Crain15,Schaye15}. A key
advantage of our approach is that we can compare our results to models
from the cosmological simulations both with and without AGN
feedback included, to allow us to identify the signature of AGN
feedback on the star forming properties of galaxies \citep[also see
  e.g.][]{Beckmann17,Harrison17}.

EAGLE is a suite of cosmological hydrodynamical simulations, which
uses an enhanced version of the GADGET-3 code \citep{Springel05} which
consists of a modified hydrodynamics solver, time-step limiter, and
employs a subgrid treatment of baryonic physics. The subgrid physics
takes into account of the stellar-mass loss, element-by-element
radiative cooling, star formation, black-hole accretion (i.e.,\ AGN
activity), and star formation and AGN feedback. The free parameters of
the subgrid physics were calibrated on the stellar mass function,
galaxy size, and the black-hole--spheroid relationships at $z \approx
0.1$ \citep{Crain15,Schaye15}. The simulation is able to reproduce a
wide range of observations of low and high redshift galaxies
\citep[e.g.,\ fraction of passive galaxies, Tully-Fisher relation,
  evolving galaxy stellar mass function, galaxy colours and the
  relationship between black hole accretion rates and SFRs; see
  e.g.][] {Furlong15,Schaye15,McAlpine17,Trayford17}. We note that, AGN feedback was introduced in the EAGLE reference model to reduce the star-formation efficiency of the most massive galaxies in order to reproduce the turn-over at the high mass end of the local galaxy stellar mass function \citep{Crain15}. The model also effectively re-produces the bi-modality of colours of local galaxies \citep[see][]{Trayford15}. However; although related, the EAGLE reference model was not directly calibrated on the parameters of the SFR or sSFR distributions at multiple epochs, making our comparison with these observables an independent test of the model.
  
In our analyses we have used two models from EAGLE: the reference
model (hereafter EAGLE ref), designed to reproduce a variety of key
observational properties (see above), and a model with no AGN feedback
(hereafter EAGLE noAGN). The EAGLE noAGN model is identical to the
EAGLE ref model in all aspects except black holes are not seeded,
which effectively turns off the AGN feedback. A comparison of the
results between these two models therefore allows for the
identification of the signature of AGN feedback on the star forming
properties of the simulated galaxies. The EAGLE ref model was run at
volumes of $25^{3}$, $50^{3}$, and $100^{3}$ cubic comoving
megaparsecs (cMpc$^{3}$). We present here the results from the largest
volume which contains the largest number of rare high-mass systems;
however, we note that we performed our analysis on all volumes and
found no significant differences in the overall results. The EAGLE
noAGN model was only performed at a volume of $50^{3}$ cubic comoving
megaparsecs. A summary of the two different EAGLE models used in our
analyses are given in Table \ref{EAGLE_tab}.

\begin{table}
 \caption{Basic properties of the EAGLE models used in the paper. From
   left to right: the model name used in the text, the reference name
   in the EAGLE database, the comoving volume (cMpc$^{3}$), the
   initial mass m$_{g}$ of the baryonic particles, and a flag to
   indicate whether AGN feedback was adopted in the model. See
   \citet{Schaye15} for more information. }  \resizebox{\columnwidth}{!}{\begin{tabular}{@{}lcccc@{}}
\hline
\hline
Model name & Database&  Volume & m$_{\rm g}$ & AGN \\
in text& Reference & (cMpc$^{3}$)& (M$_{\odot}$) & feedback?\\
\hline
EAGLE ref & RefL0100N1504 &  $100^{3}$ & $1.81 \times 10^{6}$& Yes \\
EAGLE no AGN &NoAGNL0050N0752 &  $50^{3}$& $1.81 \times 10^{6}$ & No \\
\hline
\end{tabular}}
\label{EAGLE_tab}
\end{table}

In order to construct the AGN and galaxy catalogues from the EAGLE models we
queried the public database \footnote{Available at
  http://icc.dur.ac.uk/Eagle/database.php} \citep{McAlpine16} for any
dark matter halo with a galaxy of stellar mass of $ M_{*} >2 \times
10^{10} \Msol$, for redshift snapshots over $z=$~1.4--3.6; the
slightly broader redshift range than that adopted for our main sample
ensures that the AGN and galaxy samples from EAGLE have the same mean
and median redshift as our main sample. We then applied the same
stellar mass and AGN luminosity cuts to the EAGLE sample as we used to
select our main sample. To calculate the properties of the simulated
AGN and galaxies, to allow for a systematic comparison to our main
sample, we also: 1) converted the black-hole accretion rates from the
EAGLE ref model to $\lx$ by converting them first to AGN bolometric
luminosities (assuming a nominal radiative efficiency of $\epsilon =
10\%$) and then converting to $\lx$ by multiplying it by a bolometric
correction factor of $0.1$ \citep{McAlpine17} and 2) scaled up the
SFRs calculated in both EAGLE models by $0.2$ dex to account for the
offset found by \citet{Furlong15} (see also \S 2.4 of
\citealt{McAlpine17}) from comparing the global SFR density of the
EAGLE ref model to the observed global SFR density of galaxies. Therefore, the overall galaxy population had the same selection criteria as the AGN, but we did not apply any $\lx$ threshold. The galaxies include both active and inactive galaxies as well as star-forming and passive galaxies. In
total we found $472$ AGN and $2333$ galaxies in the EAGLE ref model
and $682$ galaxies in the EAGLE noAGN model with the same properties
as in our main sample.

\begin{figure}
	\includegraphics[width=\columnwidth]{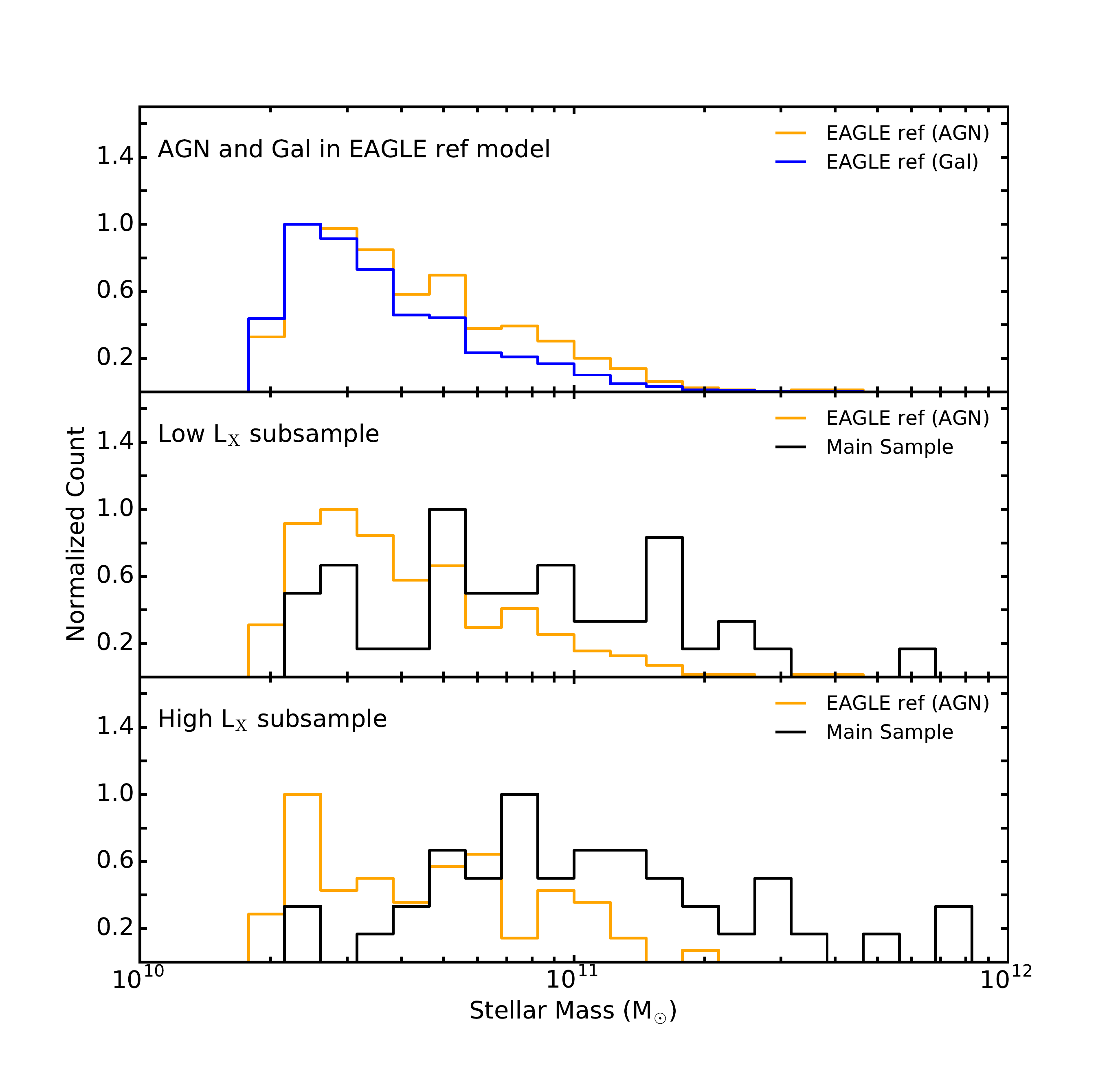}
   \caption{Comparison of the normalized stellar mass distributions
     from our different samples. Top panel: Comparison of the stellar
     mass distributions of the AGN in the EAGLE ref model (blue line)
     and galaxies in the EAGLE ref model (orange line). Middle Panel:
     Comparison of the stellar mass distribution of the low $\lx$ AGN
     in the EAGLE ref model (orange line) and the low $\lx$ AGN of the
     observed main sample (black line). Bottom panel: Comparison of the stellar
     mass distribution of the high $\lx$ AGN in the EAGLE ref model
     (orange line) with the high $\lx$ AGN of the main sample (black
     line). We take the differences in stellar mass distributions into
     consideration in \S\ref{EAGLE}.}
   \label{Figure:Mass_dist}
\end{figure}

We split the AGN in the EAGLE ref model into low and high $\lx$
subsamples using the same luminosity threshold as for our main sample
(see \S\ref{sec:sample}); the EAGLE ref low and high $\lx$ subsamples
contain 403 and 69 AGN, respectively. In Figure \ref{Figure:Mass_dist}
we compare the stellar mass distributions of the simulated AGN and
galaxies to the AGN in our main sample. The stellar mass distributions
for the AGN in the EAGLE ref model and the main sample are different
in both $\lx$ subsamples. The median stellar masses of the low and
high $\lx$ AGN in the EAGLE ref model are both $10^{10.6}$ $\Msol$. By
comparison the median stellar masses of the observed low and high
$\lx$ subsamples in our main sample are $10^{10.7}$ and $10^{11.0}$
$\Msol$, respectively. This difference in median stellar masses is
caused by the different volumes probed to select the samples. While
the EAGLE ref model has a volume of $10^{6}$ cMpc$^{3}$, the low and
high $\lx$ subsamples of our main sample were selected from larger
volumes of $10^{6.4}$ cMpc$^{3}$ and $10^{7}$ cMpc$^{3}$,
respectively.

The differences in the stellar mass distributions between the AGN in
the main sample and EAGLE will also cause the differences in the sSFR
distributions (i.e. since the sSFR distributions also depend on
stellar mass; see \S \ref{sec:modes}). We therefore have to take account of the different
stellar mass distributions to fully compare the observed and simulated
AGN. We do this using the mass matching methods described in
\S\ref{EAGLE}.

\section{Results}

In this section we present our results on the sSFR distributions of
the distant X-ray AGN in our main sample. We measure the sSFR
distributions of our main sample and search for trends in the star
forming properties as a function of $\lx$ and stellar mass (see
\S\ref{sec:modes}). To aid in the interpretation of our results we
make comparisons to the EAGLE ref model (see \S\ref{EAGLE}).

\subsection{sSFR trends with X-ray luminosity and stellar mass} \label{sec:modes}

To search for trends in the sSFR properties of the X-ray AGN, we
measured the properties (i.e.,\ the mode and the width) of the sSFR
distributions as a function of $\lx$ and stellar mass. The mode of the
sSFR distribution provides a more reliable measurement of the typical
sSFR than the linear mean (see Figure \ref{Figure:Comp_st} and
\S\ref{sec:Fits}). The width of the sSFR distribution provides a basic
measure of the range in sSFRs: a narrow width indicates that most
systems have similar sSFRs while a broad width indicates a large range
of sSFRs. We fitted log-normal distributions to the $\lx$ and stellar
mass subsamples within our main sample (see \S\ref{sec:sample}) using
the method described in \S\ref{sec:Fits}. Table~\ref{modes} presents
the overall results.

\begin{figure}
	\includegraphics[width=\columnwidth]{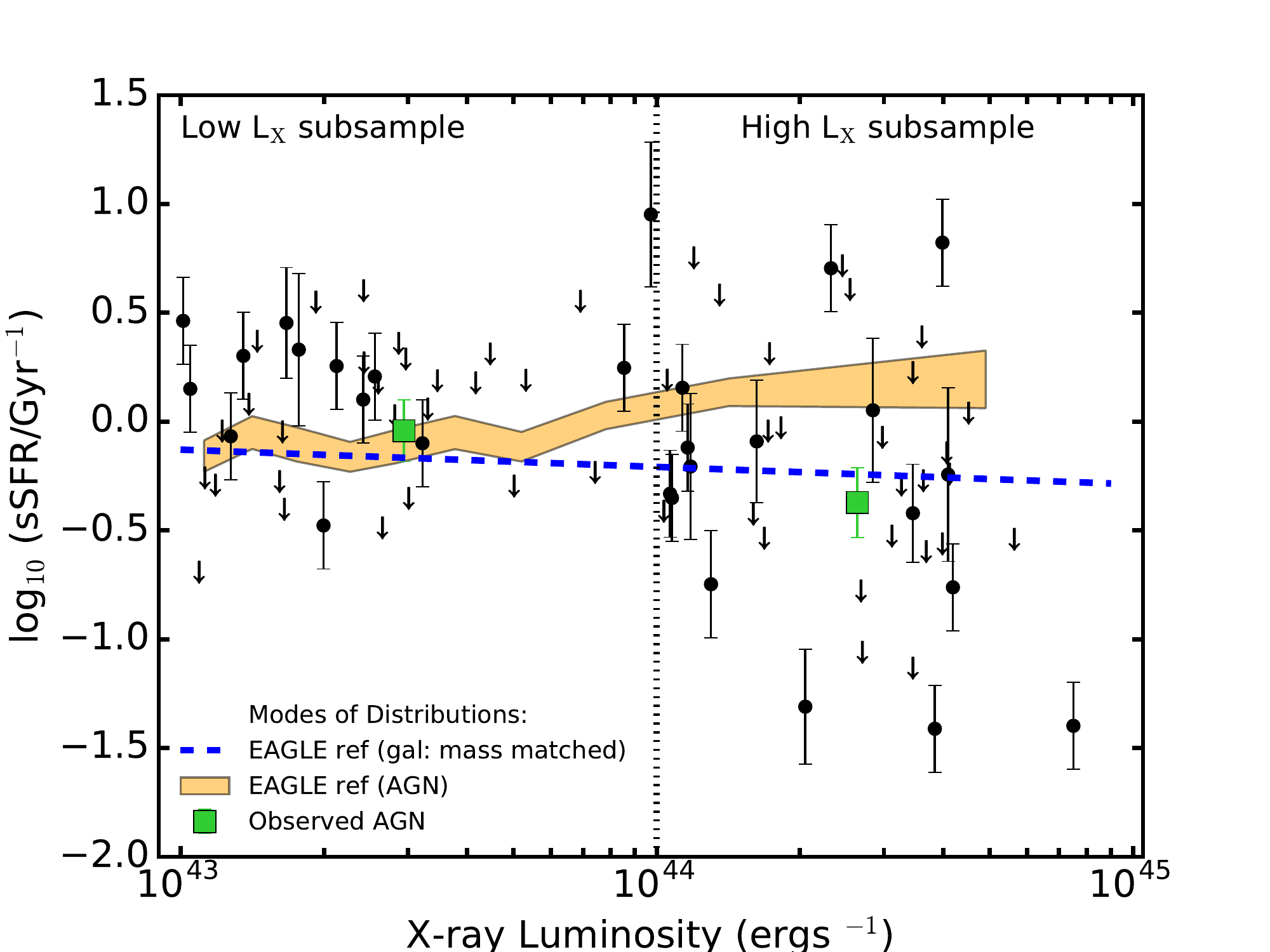}
	\includegraphics[width=\columnwidth]{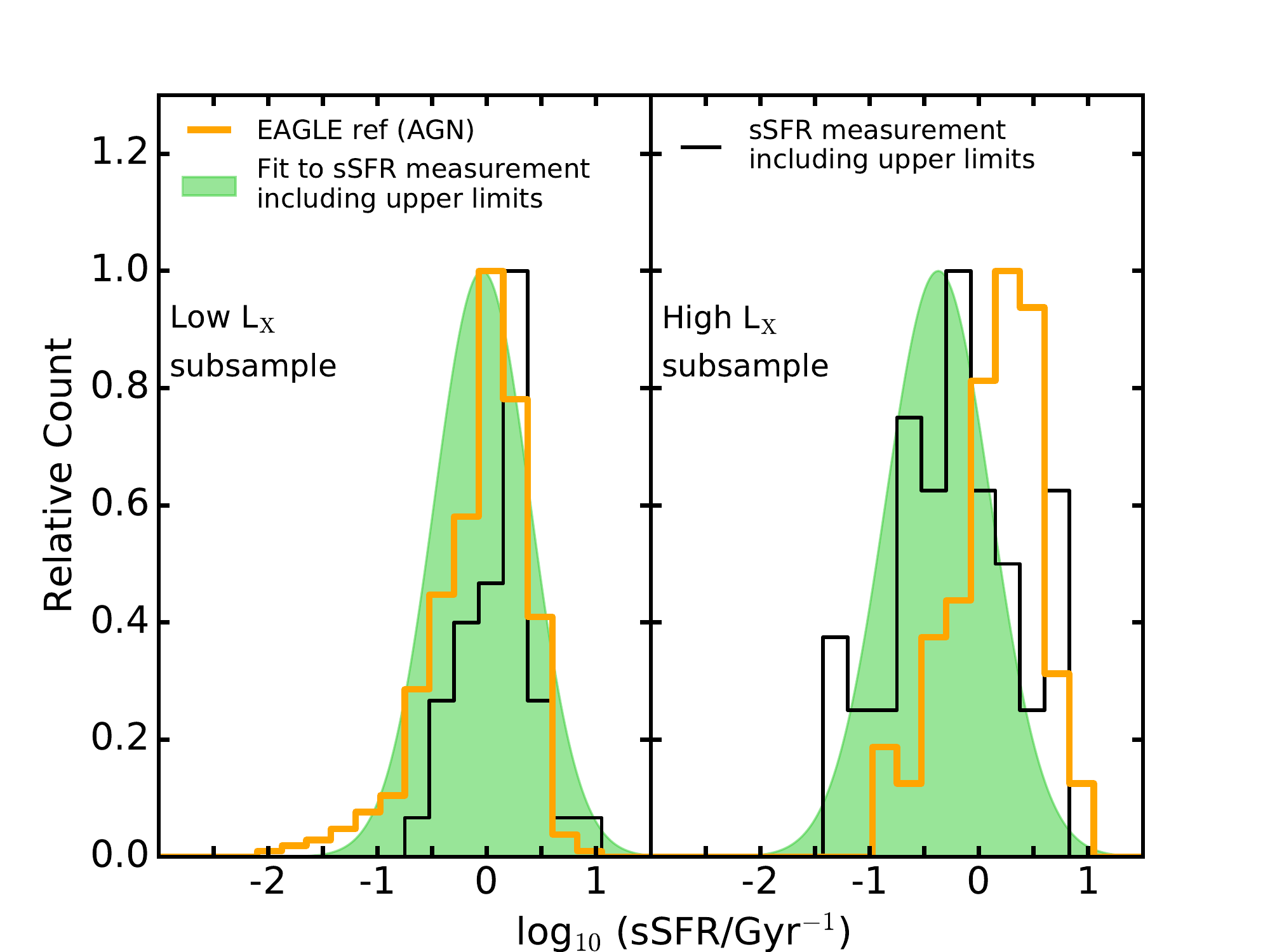}
   \caption{Top panel: sSFR versus X-ray luminosity (2--10~keV: rest
     frame) for the X-ray AGN in our main sample. The black filled
     circles indicate individual X-ray AGN, the filled green squares
     indicate the modes of the sSFR distributions for the low and high
     X-ray luminosity subsamples (see Table \ref{modes}); error bars
     represent the $68 \%$ confidence interval. The dotted vertical
     line indicates the division in X-ray luminosity between the low
     and high X-ray luminosity subsamples. The orange shaded region
     indicates the X-ray luminosity dependence on the sSFR
     distribution for AGN from the EAGLE ref model (the width
     corresponds to the $68\%$ confidence interval around the mode of
     the distribution) and the blue dashed line indicates the
     predicted sSFR--X-ray luminosity relationship from the EAGLE ref
     model for galaxies with masses matched to those found from our
     observed X-ray AGN (see \S \ref{EAGLE}). Bottom panel: sSFR
     distributions for our data (black histogram), the AGN from the
     EAGLE ref model (orange open histogram), and the best-fitting
     log-normal distribution (green filled histogram; see \S
     \ref{sec:Fits}). The sSFR distributions are shown separately for
     the low (left) and high (right) X-ray luminosity subsamples.}
   \label{Figure:LX_plot}
\end{figure}

In Figure \ref{Figure:LX_plot}, we plot the sSFR properties
(individual measurements and measurements of the distributions) of the
main sample as a function of $\lx$. The modes
(log$_{10}$($\mu$/Gyr$^{-1}$)) of the sSFR distributions of the low
$\lx$ and high $\lx$ subsamples are $0.03^{+0.14}_{-0.17}$ and
$-0.32^{+0.15}_{-0.17}$, respectively. The mode of the sSFR decreases
with $\lx$, but the drop is modest ($1.5\sigma$), ruling out a simple
AGN-feedback model where high-luminosity AGN instantaneously shut down
SF. We also note that the same qualitative result is obtained if we
consider the mean sSFR rather than the mode; however, the mean values
are $\approx$~0.3--0.5~dex higher than the mode (see
Table~\ref{modes}). The widths of the sSFR distributions for the low
$\lx$ and high $\lx$ subsamples are also consistent, with values of
$0.52^{+0.13}_{-0.10}$ and $0.65^{+0.15}_{-0.11}$, respectively.

Our results shows no evolution of the sSFR distribution with $\lx$. This general conclusion agrees qualitatively with results of most previous studies at these redshifts that investigated the mean (s)SFR as a function of $\lx$ \citep{Harrison12,Rosario12,Rovilos12,Azadi15,Stanley15,Lanzuisi17}. Here, for the first time, we have constrained the sSFR distribution properties for the AGN host galaxies at these redshifts. These results demonstrate that the previous finding of a flat trend is a true reflection of the behaviour of the typical AGN population (as measured using the mode), rather than an inaccurate description of the population. However, as expected we showed that the bulk of the population (mode) has a lower sSFR than linear mean.

\begin{figure}
	\includegraphics[width=\columnwidth]{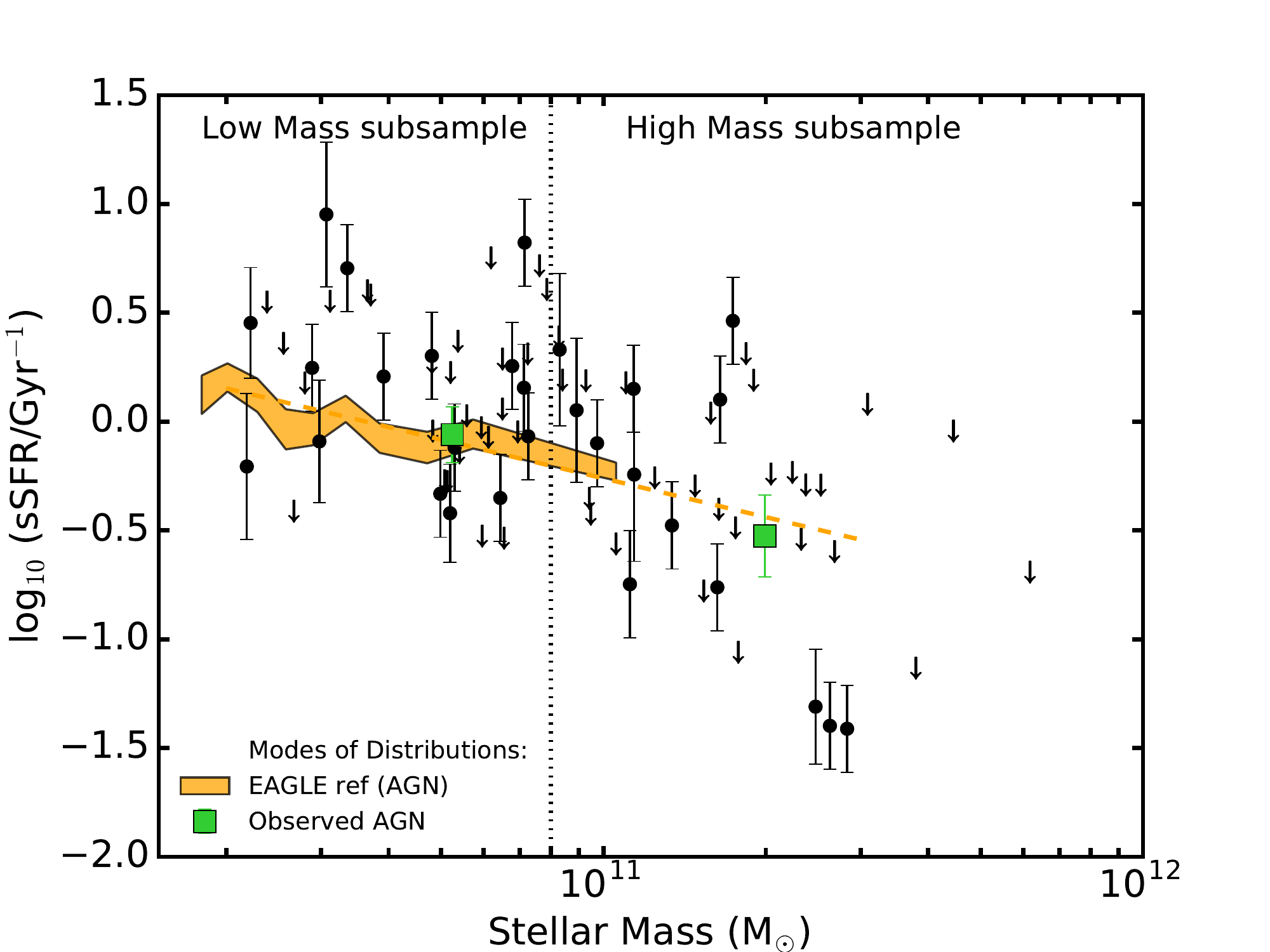}
	\includegraphics[width=\columnwidth]{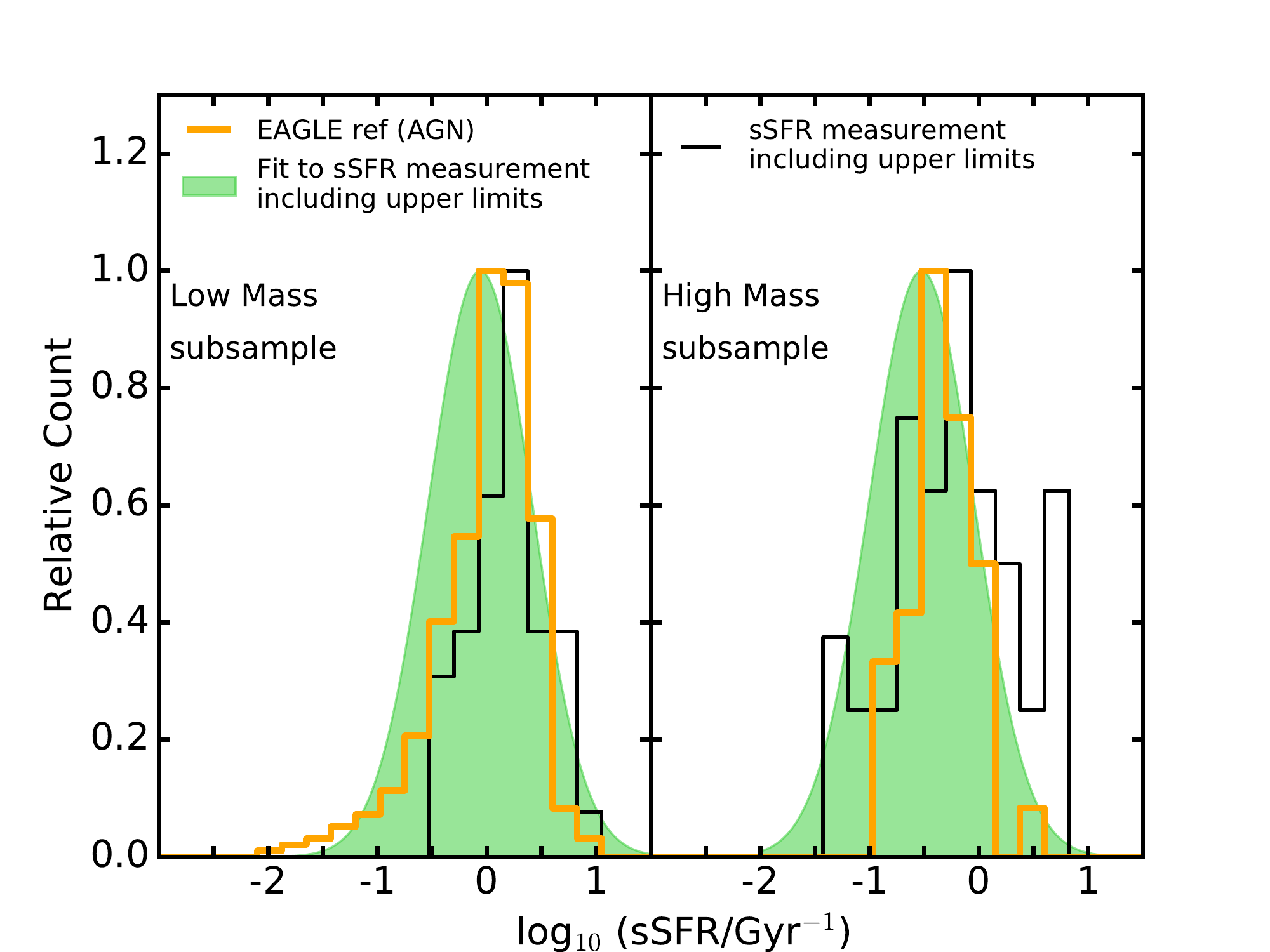}
   \caption{Top panel: sSFR versus stellar mass for the X-ray AGN in
     our main sample. The black filled circles indicate individual
     X-ray AGN, the filled green squares indicate the modes of the
     sSFR distributions for the low and high mass subsamples (see
     table \ref{modes}); the error bars represent the $68 \%$
     confidence interval. The dotted vertical line indicates the
     division in mass between the low and high stellar mass
     subsamples. The orange shaded region indicates the stellar mass
     dependence on the sSFR distribution for AGN from the EAGLE ref
     model (the width corresponds to the $68 \%$ confidence interval
     around the mode of the distribution) and the orange dashed line
     is the linear extrapolation of the mode to higher stellar masses
     (see \S \ref{sec:modes}). Bottom panel: sSFR distributions for
     our data (black histogram), the AGN from the EAGLE ref model
     (open orange histogram), and the best-fitting log-normal
     distribution (green filled histogram; see \S \ref{sec:Fits}). The
     sSFR distributions are shown separately for the low (left) and
     high (right) stellar mass subsamples.}
   \label{Figure:Mass_plot}
\end{figure}

In Figure \ref{Figure:Mass_plot}, we plot the sSFR properties
(individual measurements and measurements of the distributions) of the
main sample as a function of stellar mass. Quantitatively similar
results are obtained to those shown in Figure \ref{Figure:LX_plot} for
the sSFRs as a function of $\lx$, with no clear evidence for a strong
change in the sSFR properties towards high stellar mass: the mode
(log$_{10}$($\mu$/Gyr$^{-1}$)) and width of the sSFR distribution for
the low stellar mass subsample is $-0.01^{+0.13}_{-0.15}$ and
$0.53^{+0.13}_{-0.08}$ respectively, while the mode
(log$_{10}$($\mu$/Gyr$^{-1}$)) and width of the sSFR distribution for
the high stellar mass subsample is $-0.48^{+0.17}_{-0.20}$ and
$0.67^{+0.18}_{-0.12}$ respectively. However, the difference in the
mode of the sSFR distributions between the two stellar mass subsamples
is marginally more significant ($2.0\sigma$) than between the two
$\lx$ subsamples. Again, the mean sSFRs are also
$\approx$~0.3--0.5~dex higher than the modes (see Table~\ref{modes}). 
We put our results into context in section \ref{sec:location}.

\subsection{Comparison to the EAGLE simulations}\label{EAGLE}

The EAGLE ref model (see Table \ref{EAGLE_tab}) reproduces the global
properties of the galaxy population (see \S\ref{sec:EAGLE_d}). To help
interpret our results from \S\ref{sec:modes}, we investigate whether
the simulated AGN in this model show the same sSFR relationships as we
have found among the main sample we observed. The properties of the sSFR
distributions are calculated for the EAGLE AGN in the same $\lx$ and
stellar-mass bins as for our main sample, following \S\ref{sec:Fits};
see Table \ref{modes}. To further aid in the comparison, we also
calculated the running mode of the sSFR in $\lx$ and stellar-mass bins
of 50 objects, following \S\ref{sec:Fits}.

In Figures \ref{Figure:LX_plot} and \ref{Figure:Mass_plot}, we compare
the sSFR distributions of the EAGLE AGN to our main sample as a
function of $\lx$ and stellar mass, respectively. From these figures
and Table \ref{modes}, we note that EAGLE can generally reproduce the
widths of the observed sSFR distributions of AGN. At low $\lx$ and
stellar mass, the modes of the sSFR distributions for the EAGLE AGN
are also in good agreement with those of the main sample, but they
deviate marginally at high stellar mass, and strongly at high $\lx$.

We can qualitatively understand the marginal difference in the sSFR
modes with stellar mass (see Figure \ref{Figure:Mass_plot}) as due to
the different stellar mass distributions between the simulated AGN in
EAGLE and the observed AGN in the main sample. There are more massive
AGN hosts in the main sample than in the EAGLE ref model, which is a
consequence of the different volumes probed by the EAGLE simulation
and our observational survey (see \S\ref{sec:EAGLE_d} and Figure
\ref{Figure:Mass_dist}). Since sSFR is a decreasing function of
stellar mass, the more massive AGN in the main sample will have lower
sSFRs than the less massive AGN. Indeed, if we extrapolate the running
mode of the sSFR from the EAGLE ref model towards high stellar masses
(the dashed line in Figure \ref{Figure:Mass_plot}), we can fully
reproduce the mode of the sSFR among the observed high mass AGN hosts.

Figure \ref{Figure:Mass_dist} shows that the stellar masses of the
observed AGN and the simulated AGN from the EAGLE ref model differ
substantially in the two $\lx$ bins. This difference in stellar
mass could also be the driver of the significant differences in the sSFR
mode as a function of $\lx$ seen between EAGLE and the main sample
(see Figure \ref{Figure:LX_plot}).  We explore this idea by
considering how the mode of the sSFR changes for subsamples with
different stellar mass distributions using the EAGLE ref
model. Unfortunately, in the limited volume of the EAGLE simulation
there are no AGN hosts with masses $> 2\times 10^{11}$
M$_{\odot}$. Therefore, we turn to the more numerous galaxy population
in the EAGLE ref model. So long as the sSFRs of these simulated
galaxies decrease with stellar mass in the same functional form as the
AGN, we can use them as analogues to understand the role of differing
stellar mass distributions in the interpretation of the sSFR
differences between the simulated and observed AGN. In Figure
\ref{Figure:Ratios} we compare the mode of the sSFR distribution
versus the stellar mass for both the AGN and galaxies in the EAGLE ref
model and demonstrate that they follow the same trend but with a
$\approx$~0.1~dex offset (which we further explore in \S4.1).

To quantify the impact of different stellar mass distributions on our
results we constructed four subsets of galaxies from the EAGLE ref
model that are matched in their mass distributions to 1) simulated AGN
from the EAGLE ref model in the low $\lx$ bin, 2) simulated AGN from
the EAGLE ref model in the high $\lx$ bin, 3) observed AGN from the
main sample in the low $\lx$ bin, and 4) observed AGN from the main
sample in the high $\lx$ bin. For each of these four subsets, we
determined the mode of the sSFR distribution following the method in
\S\ref{sec:Fits}. If differences in stellar mass are the principal
driver for the different trends shown by the observed and simulated
AGN in Figure \ref{Figure:LX_plot}, we would expect offsets in the
sSFR modes of the mass-matched subsets corresponding to the simulated
and observed AGN in each respective $\lx$ bin, particularly at high
$\lx$ where the stellar mass differences are most pronounced (see
Figure \ref{Figure:Mass_dist}). This is indeed what we find.

The mode of the sSFR for the two mass-matched EAGLE galaxy subsets
corresponding to the low $\lx$ bin differ by only a small amount ($<
0.1$ dex), as expected given the similar stellar mass distributions
(see Figure \ref{Figure:Mass_dist}) and in agreement with the results
for this $\lx$ bin given in Table \ref{modes}. On the other hand, the
mode of the sSFRs for the two mass-matched EAGLE galaxy subsets
corresponding to the high $\lx$ bin differ by $\approx 0.4$ dex. From
this we conclude that the high masses of the high $\lx$ AGN in the
main sample leads to a measured sSFR that is lower than that of
equivalently X-ray luminous simulated AGN from the EAGLE ref model. If
we correct the sSFR trend with $\lx$ for the EAGLE AGN to reflect the
different stellar mass distributions of the observed AGN, using the
offsets determined above, we obtain the blue dashed line in Figure
\ref{Figure:LX_plot}, which is a remarkably good match to our
observations.

We have shown that even though EAGLE has not been calibrated on (s)SFR distributions of AGN, it reproduces accurately the shape and the parameters (mode and width) of the distribution. Furthermore, we have found that the properties of the sSFR distributions are more strongly related to stellar mass than to AGN luminosity. We investigated what these results mean in terms of AGN feedback in \S \ref{sec: NoAGN}.

\section{Discussion} \label{Discussion}

On the basis of our results on the fitted sSFR distributions of X-ray
AGN at $z=1.5-3.2$ we found that, once the effects of different
volumes and survey selections are taken into account (in particular
with respect to stellar mass distributions), the EAGLE ref model
provides a good description of the sSFR properties of the AGN in our
main sample. The good agreement between the observations and EAGLE
means that we can employ further comparisons to explore the connection
between galaxies and AGN and the role of AGN feedback in producing the
SF properties of the galaxy population.

\subsection{AGN among the galaxy population at $z\approx$~1.5-3.2}\label{sec:location}

In our study so far we have considered the star forming properties of distant
AGN but we have not put these results within the content of the
overall galaxy population. Previous studies at this redshift compare the AGN population to star-forming main sequence and over- all galaxy population. We note that our sample (Section \ref{sec:sample}) is purely an AGN and mass-selected sample and therefore potentially contains both star-foming and quiscent galaxies. Here we put our study into context with previous studies and as well as clarify the discussion in the literature. 

\begin{figure}
	\includegraphics[width=\columnwidth]{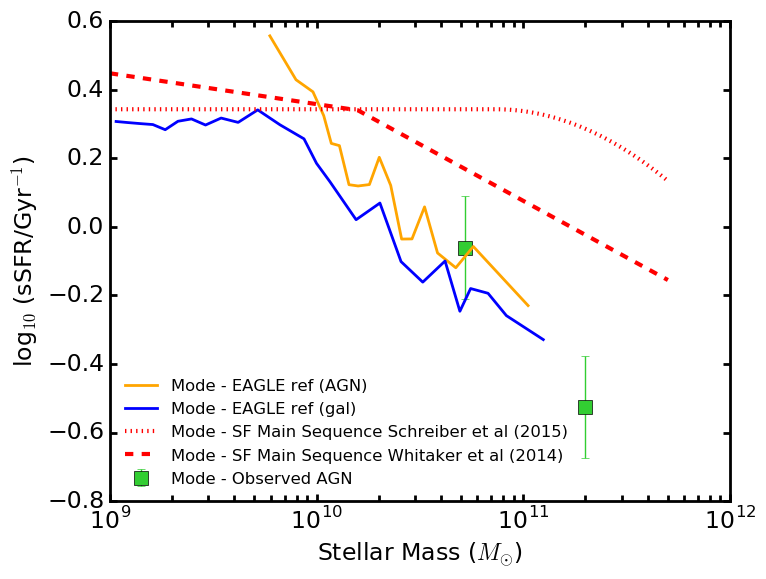}
    \includegraphics[width=\columnwidth]{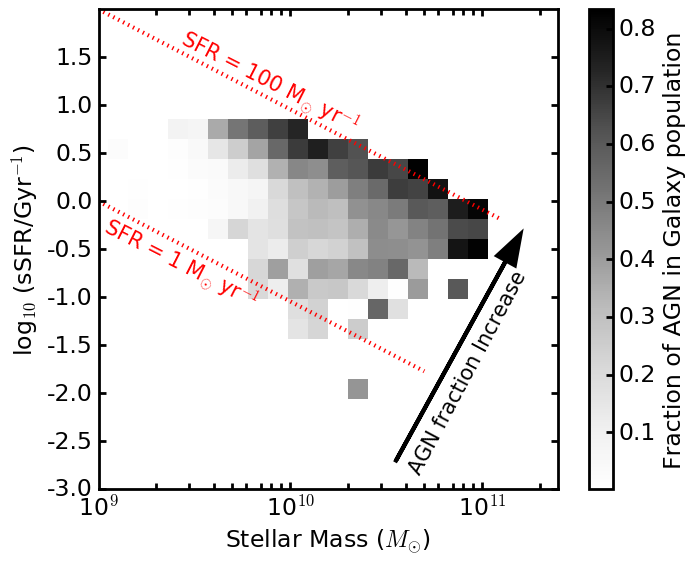}
   \caption{Top Panel: sSFR versus stellar mass for the X-ray AGN in
     our main sample and AGN and galaxies in the EAGLE ref model. The
     green filled squares indicate the mode of the sSFR distributions
     for the observed X-ray AGN with error bars representing the $68 \%$
     confidence interval (see Table \ref{modes}) and are compared to
     the modes of the AGN (orange curve) and galaxies (blue curve)
     from the EAGLE ref model, coeval ($z\approx$~2.2) main sequence
     galaxies from \citet{Schreiber15} (red dotted line) and
     \citet{Whitaker14} (red dashed line). The mode of the sSFR for
     AGN is higher than the overall galaxy population but lower than
     galaxies on the star-forming main sequence. Bottom Panel: The
     grey shaded regions indicate the fraction of galaxies in a given
     sSFR--stellar mass bin that host AGN activity (with $\lx >
     10^{43} \ergss$) in the EAGLE ref model; the AGN fraction values
     are indicated by the greyscale bar to the right of the
     figure. The dotted red lines indicate constant values of SFR. The
     fraction of galaxies hosting AGN activity in the EAGLE ref model
     is a function of the SFR (illustrated by the black arrow).}
   \label{Figure:Ratios}
\end{figure}

In Figure \ref{Figure:Ratios} we compare
the mode of the sSFR versus stellar mass for our main sample to that
of the main sequence for coeval star-forming galaxies.\footnote{We
  used the parameters from Table 1 of \citet{Mullaney15} to convert
  between the linear mean and the mode of the sSFR distribution of the
  star-forming galaxy main sequence.}
Although there is some uncertainty in the sSFR of the main sequence at this redshift and high mass, the AGN clearly lie substantially ($\approx$~0.2--0.8~dex) below it, particularly at higher stellar mass (see dotted and dashed tracks in Figure \ref{Figure:Ratios}). The top panel of Figure \ref{Figure:Ratios} is in good agreement with earlier studies and demonstrates that a fraction of the X-ray AGN population (equivalent to the orange line) do not lie in star-forming galaxies \citep[red dashed and dotted lines;][]{Nandra07, Hickox09, Koss11, Mullaney15}, even though Herschel-based studies suggest that they are more star-forming on \textit{average} than the overall galaxy population \citep[equivalent to blue line; also see ][]{Santini12,Rosario13b,Vito14b, Azadi17}. This is also found for local (z=0) X-ray AGN \citep{Shimizu15}.

Given the good agreement between our observational results and the
EAGLE ref model (see \S\ref{EAGLE}), we can use EAGLE to provide
additional insight on the connection between distant galaxies and
AGN. In Figure \ref{Figure:Ratios} (top panel) we show that the sSFR
properties of the AGN in EAGLE are $\approx$~0.1~dex higher than the
galaxies in EAGLE, at a given stellar mass. This indicates that,
although AGN do not typically reside in strong star-forming galaxies,
their SFRs are elevated when compared to the overall galaxy
population. In Figure \ref{Figure:Ratios} (bottom panel) we show the
fraction of galaxies that host an AGN with $\lx > 10^{43} \ergss$ in
the EAGLE ref model across the sSFR--stellar mass plane. The fraction
of galaxies hosting an AGN increases as a function of both sSFR and
stellar mass (i.e.,\ effectively as a function of SFR), from an AGN
fraction of $<10$\% at low values to $>50$\% at high SFR values 
(SFR >$ 50 \Msolyr$). Overall the highest AGN fractions are found for galaxies with
the highest SFRs, suggesting a connection between the cold-gas supply required to
fuel intense star formation and the gas required to drive significant
AGN activity \citep{Silverman09}. By selecting AGN with 
$\lx > 10^{43} \ergss$ we are
therefore biased towards galaxies with elevated SFRs when compared to
the overall galaxy population. This effect is responsible for the
$\approx$~0.1 -0.2~dex difference in the sSFR properties between galaxies
and AGN in the EAGLE ref model (see Figure \ref{Figure:Ratios}).

\subsection{Identifying the signature of AGN feedback on the star forming properties of galaxies}\label{sec: NoAGN}

Our analyses of the EAGLE simulation in \S\ref{sec:location} suggested that AGN 
have elevated sSFRs when compared to the overall galaxy population. Furthermore, 
both the data and the model do not reveal a negative trend between sSFR and 
AGN luminosity (see Figure \ref{Figure:LX_plot}). These results may
appear counter intuitive for a model in which AGN feedback quenches
star formation in galaxies. Therefore, what is the signature of AGN feedback on
the star-forming properties of galaxies? This question can be explored from a
comparison of the sSFR properties of galaxies and AGN for two
different EAGLE models: the EAGLE ref model with AGN feedback and the
EAGLE noAGN model, which is identical to that of the EAGLE ref model
except that black holes are not seeded in this model and consequently
there is no AGN activity and no AGN feedback (see
\S\ref{sec:EAGLE_d}).

\begin{figure*}
	\includegraphics[width=0.6\paperwidth]{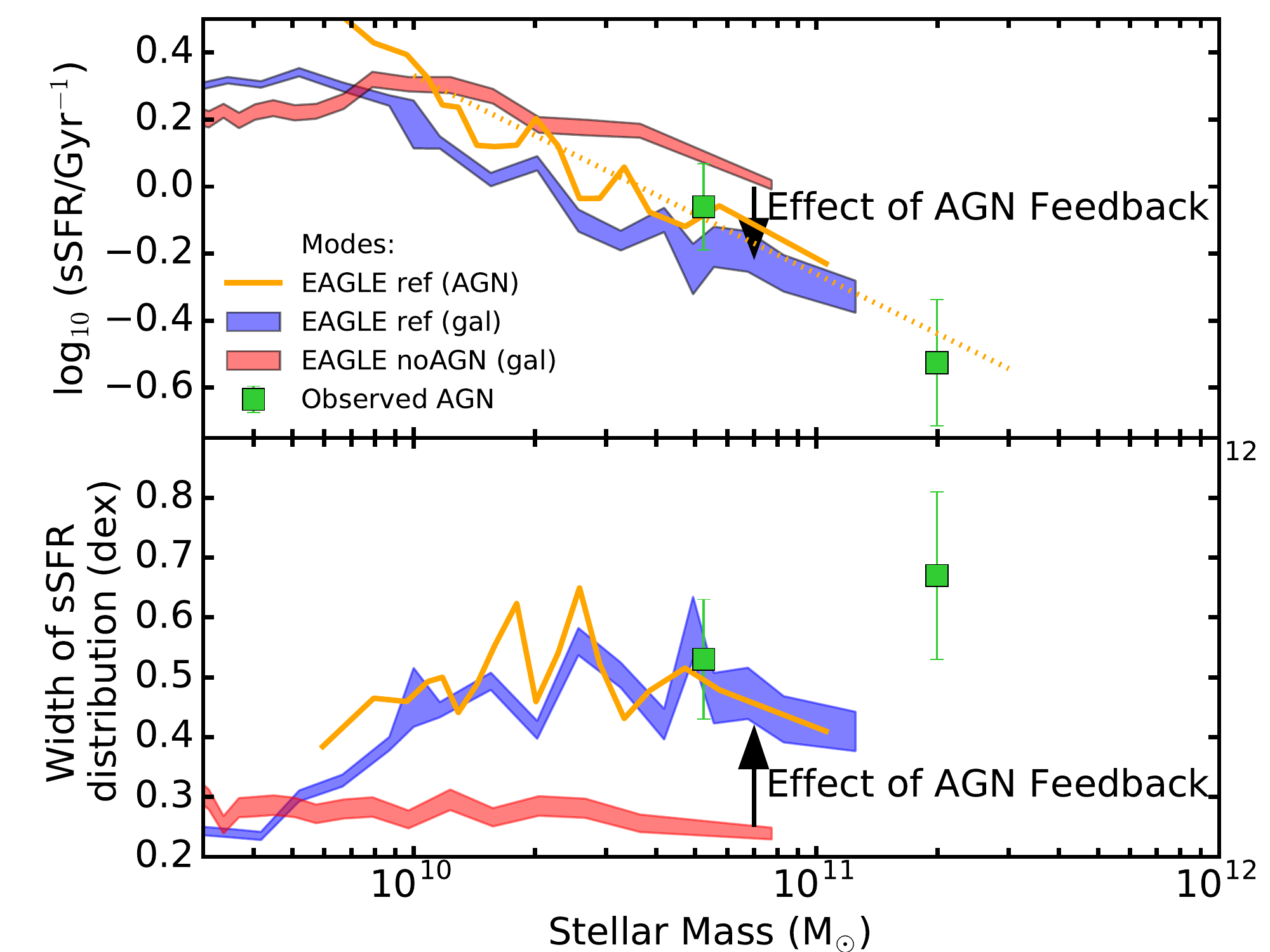}
   \caption{Mode of the sSFR (top panel) and width of the sSFR (bottom
     panel) versus stellar mass for the X-ray AGN in our main sample
     and two different EAGLE models. The solid green squares indicate
     the measurements from the X-ray AGN in our main sample; the error
     bars indicate the $68 \%$ confidence interval (see Table
     \ref{modes}). The blue and red shaded regions indicate the modes
     and widths of the sSFR for galaxies in the EAGLE ref model and
     the EAGLE model without AGN, respectively. The orange solid line
     indicates the modes and widths of the sSFR for AGN in the EAGLE
     ref model and the orange dashed line in the top panel indicates
     the linear extrapolation to higher stellar masses.}
   \label{Figure:Money_plot}
\end{figure*}

We calculated the running mode and width of the sSFR distributions for
the galaxies in both the EAGLE ref model and the EAGLE noAGN model in
stellar-mass bins of 50 objects, following \S\ref{sec:Fits}. In Figure
\ref{Figure:Money_plot} we compare the mode and width of the sSFR
distributions of the galaxies between these two models. There are
several clear differences between the sSFR properties of the galaxies
with $>10^{10} \Msol$ in the EAGLE ref and the EAGLE noAGN models: 1)
the sSFR distribution is a factor $\approx$~2 broader in the EAGLE ref
model, 2) the mode of the sSFR is $\approx$~0.2~dex lower in the EAGLE
ref model, and 3) the slope of the mode of sSFR distribution as a
function of mass is steeper in the EAGLE ref model;
$-0.52 \pm 0.02$ and $-0.35
\pm 0.02$ for the EAGLE ref and EAGLE noAGN model, respectively, when 
we fitted a linear model to the data in logarithmic space. Of
these three potential signatures of AGN feedback, we consider the
broadening of the sSFR distribution to be the most reliable quantity
for comparison with observations since it is less sensitive to
calibration differences in stellar mass and SFR calculations between
the observations and simulations. Furthermore, the width of the sSFR distributions is more sensitive to the effect of AGN feedback, since it is sensitive to a decrease in the sSFR for even a small fraction of the population.

In Figure \ref{Figure:Money_plot} we
compare the sSFR properties of the AGN in the EAGLE ref model to the
galaxies in the same model. These signatures of AGN feedback are seen
in both the AGN and galaxy population, implying that the impact of AGN
feedback is slow and occurs on a timescale that is longer than the
episodes of AGN activity \citep[see][]{Harrison17, McAlpine17}. This slow impact of AGN
feedback on the star forming properties helps to explain why AGN luminosity
($\lx$) is not observed in the data for the EAGLE reference model to be a strong 
driver of the sSFR properties (see Figure 4); i.e.,\ although the luminosity of
the AGN may dictate the overall impact of the feedback on star formation, the
observational signature of that impact on the star formation across the galaxy
is not instantaneous. However, we note that since the measurements of star 
formation in our study are for the entire galaxy, these results do not rule out 
AGN having significant impact on a short timescale on the star formation in localised regions within
the galaxy. Also the fact that the signature of AGN feedback is in both the AGN and the overall galaxy population implies that we do not have to solely study the AGN in order to understand the AGN feedback, i.e. constraining the sSFR distribution of overall galaxy population can help determine the effect of AGN feedback on star formation.

\begin{figure}
	\includegraphics[width=\columnwidth]{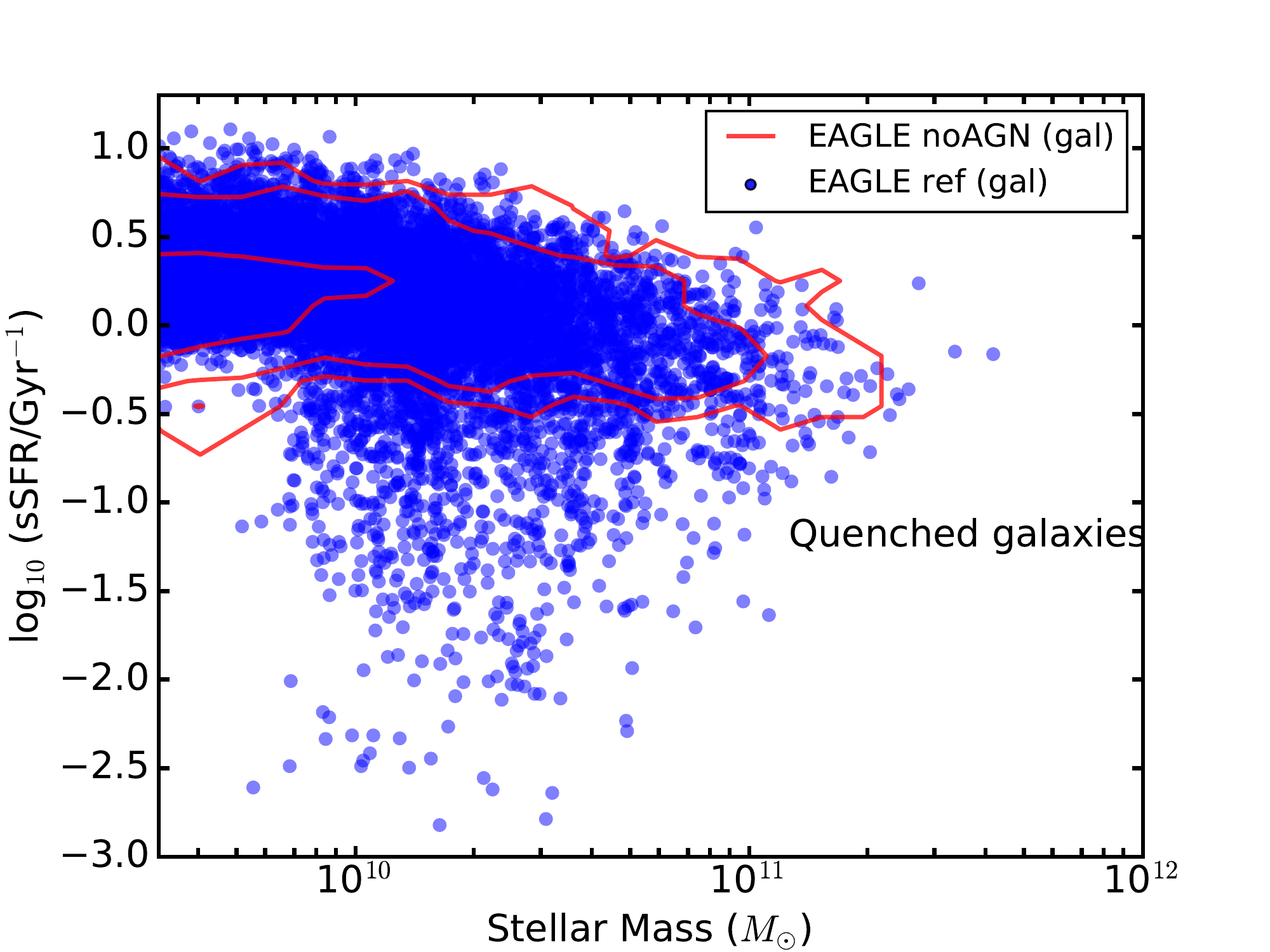}
   \caption{Individual galaxies from the EAGLE ref (blue points) and
     1, 2 and 3 $\sigma$ contours (red lines) of the galaxies in the
     EAGLE noAGN model on the sSFR--stellar mass plane. In the EAGLE
     noAGN model, there are no galaxies with
     $\log_{10}$(sSFR/Gyr$^{-1}$)<$-0.5$ Gyr$^{-1}$. The sSFR
     distributions in the EAGLE ref model is a factor $\approx$~2
     broader than in the EAGLE noAGN model.}
   \label{Figure:Contours}
\end{figure}

In Figure \ref{Figure:Money_plot} we show how the measured sSFR
properties of the AGN in our main sample compare to systems in the
EAGLE ref and noAGN models. From this comparison it is clear that the
broad width of the sSFR distribution for our main sample is in better
agreement with the EAGLE ref model than the EAGLE noAGN model,
providing indirect observational support for the AGN feedback in EAGLE. The broad
width of the sSFR distribution indicates a wide range in sSFRs. This
is seen in Figure \ref{Figure:Contours}, where we compare the sSFR
versus stellar mass for the galaxies in the EAGLE ref and the EAGLE
noAGN models. The clearest differences between the two models across
the sSFR--stellar mass plane are the broader range of sSFRs for the
galaxies in the EAGLE ref model and the presence of a population of
galaxies with low sSFRs (less than $\log_{10}$(sSFR/Gyr$^{-1}$)=$-0.5$
Gyr$^{-1}$) not seen in the EAGLE noAGN model.

Since the two EAGLE models are identical except for the presence/absence of AGN feedback, perhaps unsurprisingly, we conclude that AGN are primarily responsible for creating the low sSFR (``quenched'') part of the galaxy population in the EAGLE ref model \citep{Trayford16}. The halo mass quenching which is present in both models is partially responsible for a small decrease of sSFR with stellar mass, but does not reproduce the observed width and mode of the sSFR distributions (see Figure \ref{Figure:Money_plot}). Importantly, the EAGLE ref model was not calibrated to reproduce the properties of (s)SFR distributions at any redshift but successfully reproduces the parameters we measured from our observations. We have shown that we would not expect to see a strong signature of AGN feedback in trends of sSFRs as a function of AGN luminosity, but instead in the reduced mode and increased width of the sSFR distributions for the most massive galaxies.

\section{Conclusions}

We observed 114 X-ray selected AGN with ALMA at $870 \mu$m across a
broad range in luminosity ($\lx =5 \times 10^{39}-10^{45} \ergss$) and
redshift ($z=0.1-4.6$). Utilising the ALMA data in combination with
archival \textit{Herschel} and \textit{Spitzer} data, we fitted the
broad-band SEDs to obtain SFR and stellar-mass measurements
uncontaminated by AGN emission. In the current paper we focused our
analyses on a main sample of 81 X-ray selected AGN (irrespective of
ALMA coverage) at $z=1.5-3.2$ with $\lx =10^{43}-10^{45} \ergss$ and
stellar mass of $>2 \times10^{10} \Msol$. We used the SFR and
stellar-mass measurements to parameterise the sSFR distributions as a
function of X-ray luminosity and stellar mass, taking into account of
both detections and upper limits using Bayesian techniques. To assist
in the interpretation of our results, we made comparisons to the
predictions from two different models from the EAGLE hydrodynamical
cosmological simulation: the reference model (EAGLE ref model), which
includes AGN feedback, and a model without black holes which,
consequently, does not include AGN feedback (EAGLE noAGN). On the
basis of our analyses we obtained the following results:

\begin{enumerate}

\item We found no strong ($>3$~$\sigma$) observational evidence for
  differences in the mode or width of the sSFR distribution for the
  AGN in our main sample as a function of $\lx$. The lack of a 
  dependence on the sSFR properties with $\lx$
  rules out a simple AGN-feedback model where high-luminosity AGN
  instantaneously shut down star formation. However, we do find good
  agreement between the properties of the sSFR distributions of our
  main sample and the EAGLE ref model as a function of both $\lx$ and
  stellar mass, although only when the samples are matched in
  mass. This result indicates the importance of taking account of
  stellar mass in sSFR comparisons. See \S\ref{sec:modes} and
  \S\ref{EAGLE}.

\item From a comparison of the properties of the sSFR distributions of
  the galaxies in the EAGLE ref model to the galaxies in the EAGLE
  noAGN model we identified a clear signature of AGN feedback on the
  star forming properties of galaxies. We found that the sSFR
  distribution is significantly broader (by a factor of $\approx$~2)
  for the galaxies in the EAGLE ref model \textbf{above $2 \times 10^{10} \Msol$} due to the presence of a
  significant population of ``quenched'' galaxies with low sSFRs. The
  broad width of the sSFR distribution of the observed population is in better
  agreement with the EAGLE ref model than the EAGLE nonAGN model,
  providing indirect evidence for AGN feedback. See
  \S\ref{sec:location} and \S\ref{sec: NoAGN}.

\end{enumerate}

Overall, from the combination of the observations with the model
predictions, we conclude that (1) even with AGN feedback, there is no
strong relationship between the sSFR distribution parameters and
instantaneous AGN luminosity, indicating that the impact of AGN
feedback on star formation is slow and (2) a signature of AGN feedback
is a broad distribution of sSFRs for all galaxies regardless of whether they
host a AGN or not, with M$_{*}$> $10^{10}$M$_{\odot}$, which implies the
presence of a population of ``quenched'' galaxies with low sSFRs. With
future larger samples of AGN and galaxies with sensitive sSFR
measurements (e.g.,\ from deeper ALMA observations and other SFR
tracers) we aim to measure the sSFR distribution parameters of all galaxies to greater accuracy to further constrain the role of AGN in models of galaxy
formation.

\section*{Acknowledgements}

We thank the referee for constructive feedback which led to improving this work. 
We gratefully acknowledge support from the Science and Technology
Facilities Council (JS through ST/N50404X/1; DMA, CMH and DR through
grant ST/L00075X/1; TT, SM and RGB through ST/L00075X/1, ST/P000451/1, ST/K003267/1)
and the Faculty of Science Durham Doctoral
Scholarship (FS). This paper makes use of ALMA data: ADS/JAO.ALMA\#
2012.1.00869.S and ADS/JAO.ALMA\# 2013.1.00884.S. ALMA is a
partnership of ESO (representing its member states), NSF (USA) and
NINS (Japan), together with NRC (Canada) and NSC and ASIAA (Taiwan),
in cooperation with the Republic of Chile. The Joint ALMA Observatory
is operated by ESO, AUI/NRAO and NAOJ.

This work used the DiRAC Data Centric system at Durham University, operated by
the Institute for Computational Cosmology on behalf of the STFC DiRAC HPC
Facility (\url{http://www.dirac.ac.uk}). This equipment was funded by BIS
National E-infrastructure capital grant ST/K00042X/1, STFC capital grant
ST/H008519/1, and STFC DiRAC Operations grant ST/K003267/1 and Durham
University. DiRAC is part of the National E-Infrastructure. We acknowledge
PRACE for awarding us access to the Curie machine based in France at TGCC, CEA,
Bruy\`{e}res-le-Ch\^{a}tel.

%%%%%%%%%%%%%%%%%%%%%%%%%%%%%%%%%%%%%%%%%%%%%%%%%%



\bibliographystyle{mnras}
%\bibliography{mybib} % if your bibtex file is called mybib.tex

%%%%%%%%%%%%%%%%%%%%%%%%%%%%%%%%%%%%%%%%%%%%%%%%%%

%%%%%%%%%%%%%%%%% APPENDICES %%%%%%%%%%%%%%%%%%%%%

\appendix

\section{ALMA observations and catalogues}

In this appendix we describe the band~7 (870~$\mu$m) ALMA observations
and the construction of the ALMA catalogues for the X-ray AGN observed
from our cycle 1 (project 2012.1.00869.S; PI: J.~Mullaney) and cycle 2
(project 2013.1.00884.S; PI: D.~Alexander) programmes. A subset of the
ALMA-observed X-ray AGN are used in our main analyses, as described in
\S2, and SFR constraints for all of the ALMA-observed X-ray AGN at
$z>1$ are presented in Stanley et~al. (in prep); we note here that the
SFRs in Stanley et~al. (in prep) can differ by up-to $0.1$~dex from
those presented here due to a slightly different method adopted to
select the best-fitting SED solution (see \S \ref{sec:SED}).

Here we provide an overview of the ALMA target selection (see
\S\ref{sec:asel}), the details of the ALMA observations (see
\S\ref{sec:ALMA_obs}), the reduction of the ALMA data (see
\S\ref{sec:ALMA_DR}), the detection of ALMA sources and the matching
of ALMA-detected sources to X-ray AGN, including ALMA upper limits for
the X-ray AGN that are undetected by ALMA (see
\S\ref{sec:ALMA_props}).

\subsection{ALMA target selection}\label{sec:asel}

All of the ALMA-selected targets from our Cycle 1 and Cycle 2
programmes are X-ray AGN that are detected in either the 4~Ms
\textit{Chandra} Deep Field South (CDF-S; \citealt{Xue11}) or the
\textit{Chandra} Cosmic Evolution Survey (COSMOS) surveys
(\citealt{Civano08,Elvis09}). The overall target selection criteria
were X-ray AGN at $z>1.5$ with $\lx > 10^{42} \ergss$, for the reasons
outlined in \S\ref{sec:sample}; however, we also note that the lower
limit on the redshift selection was also required to make the most
efficient use of ALMA for SFR constraints since the sensitivity of
\textit{Herschel} for measuring SFRs is comparable to, or better than,
ALMA at $870 \mu$m for sources at $z<1.5$ (see \citealt{Casey14} for a
general review).

For the X-ray AGN in CDF-S we selected sources across the whole of the
\textit{Chandra}-observed region while for COSMOS we selected sources
from the central $12.5^{\prime}$-radius region for X-ray AGN with $\lx
= (1 - 3)\times10^{44} \ergss$ and from the central
$25^{\prime}$-radius region for X-ray AGN with $\lx = (0.3 -
1)\times10^{45} \ergss$; the larger region for the AGN with $\lx =
(0.3 - 1)\times10^{45} \ergss$ was required to allow for a comparable
number of AGN as that in the $\lx = (1 - 3)\times10^{44} \ergss$
bin. IR-based star forming luminosity constraints were obtained for
all of the X-ray AGN in CDF-S and COSMOS that met these criteria from
fitting the \textit{Spitzer}--\textit{Herschel} IR SEDs with AGN and
star forming templates, following \cite{Stanley15}. These star
formation luminosity constraints were used to select X-ray AGN to
observe with ALMA, with the majority of the selected targets having
star formation luminosity upper limits.

Overall we selected 30 X-ray AGN in CDF-S to observe in Cycle 1 and 86
X-ray AGN in CDF-S and COSMOS to observe in Cycle 2 for 116 targets
overall. The X-ray AGN selected for the Cycle 1 observations had
redshifts of $z=$~1.5--4.0 and the majority had X-ray luminosities of
$\lx \approx 10^{42} - 10^{44} \ergss$, with a minority at $\lx >
10^{44} \ergss$. The X-ray AGN selected for the Cycle 2 observations
were typically more luminous than in Cycle 1 ($\lx \approx 10^{43} -
10^{45} \ergss$) and covered the narrower redshift range of
$z=$~1.5--3.2.\footnote{We note that in selecting X-ray AGN targets
  and planning for the ALMA observations we used the redshifts, X-ray
  luminosities, and optical positions from \cite{Xue11} and
  \cite{Civano08}. However, for our analyses in this paper we have
  adopted the updated redshifts, X-ray luminosities, and optical
  positions from \cite{Hsu14} and \cite{Marchesi16}.}

\subsection{ALMA observations}\label{sec:ALMA_obs}

From the 116 X-ray AGN that we proposed for ALMA observations in cycle
1 and cycle 2 (see \S\ref{sec:asel}), 107 were observed; the 9 X-ray
AGN not observed were Cycle 2 targets in the CDF-S at
$z=$~1.5--2.0. The 107 X-ray AGN were observed by ALMA in band 7 using
a fixed continuum correlated setup with $7.5$~GHz of bandwidth
centered at $344$~GHz (870~$\mu$m) and four 128-channel
dual-polarisation basebands. The ALMA pointings were centered on the
optical counterpart positions of the X-ray sources. The Cycle 1 data
for project 2012.1.00869.S were taken on 2013 November 2 and 2013
November 16--17 using thirty-two 12~m antennas and nine 7~m antennas
in the compact array (see also \citealt{Mullaney15} for details). The
Cycle 2 data for project 2013.1.00884.S were taken on 2014 September
2, 2014 December 31, and 2015 January 1--2 using thirty-four 12~m
antennas and nine 7~m antennas in the compact array.

The requested spatial resolution for both programmes was
$\approx1^{\prime\prime}$ to ensure that the measured 870~$\mu$m
continuum emission was from the entire galaxy (physical scales of
$\approx$~7.0--8.5~kpc over the redshift range of $z=$~1.5--4.0 for
our assumed cosmology) to remove the need to apply aperture-correction
factors to match the lower-resolution \textit{Spitzer} and
\textit{Herschel} infrared data. However, the ALMA observations were
taken with a variety of baselines across both programmes (91--393~m),
which leads to some variation in the spatial resolution
($0^{\prime\prime}.18$-$0^{\prime\prime}.85$); see Tables
\ref{Flux_table} \& \ref{Flux_tableb} for the measured median baseline
for each target.

The requested sensitivity for each target was broadly based on that
required to detect star-formation emission from systems that lie on or
below the star-forming galaxy main sequence
\citep[e.g.][]{Schreiber15,Whitaker14}.  For the Cycle 1 programme the
sensitivity limits were determined taking account of both the stellar
mass and redshift of each X-ray AGN \citep[see][for more
  details]{Mullaney15} for more while for the cycle 2 programme only
the redshift was taken into account. On the basis of these parameters,
the proposed root mean squared (RMS) sensitivities varied over
0.075--0.24~mJy. However, the final sensitivities often deviated from
the proposed sensitivities due to either non-optimal conditions or
baseline configurations (i.e.,\ a more extended array configuration
than proposed). The final RMS sensitivities were re-measured from the
tapered images (see \S\ref{sec:ALMA_DR}); the final RMS sensitivities
measured for each target are given in Tables \ref{Flux_table} \&
\ref{Flux_tableb}.

\subsection{ALMA data reduction}\label{sec:ALMA_DR}

Our data reduction and source detection approach follows that
described in \citet{Simpson15}. Here we provide a brief description of
the procedures.

The data were imaged using the Common Astronomy Software Application
(CASA version 4.4.0). The uv-visibilities were Fourier transformed to
create ``dirty'' images. These dirty images were consequently
``cleaned'' using a similar technique to that described by
\citet{Hodge13}; cleaning is a common technique applied to
interferometric data to reduce the strength of the side lobes from
bright sources to allow for the detection of faint sources. We used an
iterative approach to cleaning the images. We estimated the RMS in the
dirty maps and we cleaned the maps to 3~$\sigma$ (i.e.,\ until peaks
down to 3~$\sigma$ become identifiable). We then estimated the RMS in
the cleaned maps and identified any objects at
$\geqslant$~5~$\sigma$. If a source was detected at
$\geqslant$~5~$\sigma$ then the cleaning process was repeated on the
cleaned map in a tight region around the detected source. If a source
was not detected at $\geqslant$~5~$\sigma$ then the cleaned map was
adopted as the final map.

To ensure that the 870~$\mu$m emission is measured over a common
physical size scale for all of the targets, we ``tapered'' all of the
images to give a synthesized beam of $0^{\prime\prime}.8$; this size
scale was chosen to provide 870~$\mu$m constraints from the entire
galaxy to allow for consistent comparisons with the lower-resolution
\textit{Spitzer}--\textit{Herschel} data. We applied a Gaussian taper
which lowers the weighting given to the long baselines to increase the
size of the synthesised beam. However, this procedure also increases
the noise of the maps by up-to a factor of $\approx$~6 for the
highest-resolution data. All final maps and all measured 870~$\mu$m
properties have the same spatial resolution of $0^{\prime\prime}.8$.

\subsection{ALMA source detection and source properties}\label{sec:ALMA_props}

The final maps described in \S\ref{sec:ALMA_DR} were used to detect
ALMA sources. To construct a catalogue of ALMA-detected sources we
require a clear detection threshold to reliably distinguish between
spurious sources and real detections. To provide an assessment of the
rate of spurious sources as a function of detection threshold, we
created inverted maps by multiplying the final maps by $-1$. These
inverted maps have the same noise properties as the original maps but
they do not contain any positive peaks due to real sources (all real
sources will have negative peaks).

To estimate the number of spurious sources in our final maps we
compared the ratio of sources ``detected'' in both the final maps and
inverse maps as a function of the detection threshold. To achieve this
we extracted all positive peaks of at least 2.5~$\sigma$ from the
cleaned maps corrected for the primary beam, and the inverted maps
using \textit{Source Extractor} \citep{Sextractor}. Since we are only
interested here in the ALMA properties of X-ray sources, rather than
performing a blind search for ALMA sources, our total source-detection
region size is substantially smaller than the combined area for all of
the ALMA images. Consequently, we can detect sources down to lower
significance levels than would be possible from a blind
source-detection approach. We therefore split the number of detected
peaks in the final and inverse ALMA maps into three different $\sigma$
bins: $2.5-3$ (low-significance peaks), $3-4$ (medium-significance
peaks) and $>4$ (high-significance peaks). Adopting a search radius of
$0.5^{\prime\prime}$, we calculate a total of $2.41$, $0.89$ and
$0.052$ spurious objects for the $\sigma$ bins of 2.5--3, 3--4, and
$>4.0$, respectively. Since the spurious fraction for the
high-significance bin was so small, we increased the search radius of
this bin to $1^{\prime\prime}$, which still gives a low $0.20$
spurious sources.

\begin{figure}
	\includegraphics[width=\columnwidth]{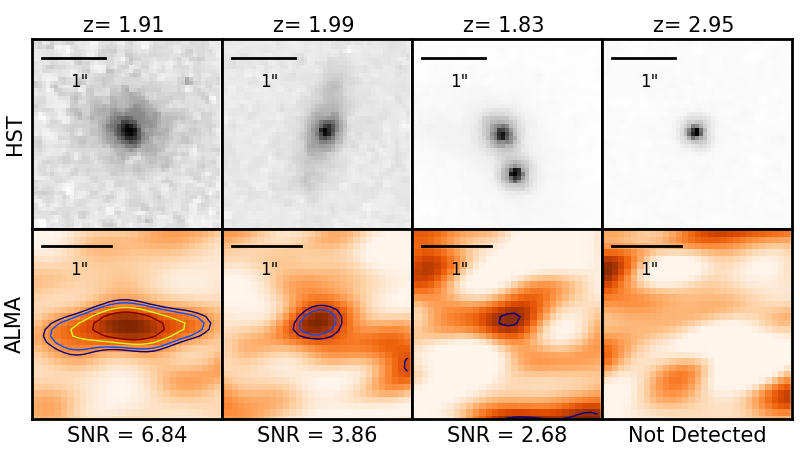}
   \caption{Example HST (H-band: $1.6 \mu$m; top) and ALMA ($870
     \mu$m; bottom) images of X-ray AGN to indicate the range in
     $\sigma$ (SNR) from our ALMA data. All images are $3^{\prime
       \prime} \times 3^{\prime \prime}$ in size; the solid bar
     indicates $1^{\prime\prime}$, which corresponds to $\approx$~8
     kpc over the redshift range for our main sample. The plotted
     contours indicate the 2.5, 3.0, 4.0, and 5.0~$\sigma$ levels for
     the ALMA data.}
   \label{Figure:Stamps}
\end{figure}

In matching ALMA sources to X-ray sources we therefore adopted a
$0^{\prime\prime}.5$ radius for low and medium significance ALMA
sources and a $1^{\prime\prime}$ radius for high-significance ALMA
sources. With this source-matching approach we identified ALMA
counterparts with a $\sigma\geqslant 2.5$ ALMA detection for 20 X-ray
sources in CDF-S and 20 X-ray sources in COSMOS.\footnote{During the
  inspection of the optical and ALMA images, we noticed a systematic
  offset between the ALMA and optical-based astrometry in the central
  GOODS-S region of CDFS ($+0.19^{\prime\prime}$ in RA and
  $-0.23^{\prime\prime}$ in declination), which was not present
  between the VLA radio data and ALMA. As noted in other papers
  (e.g.,\ \citealt{Miller08,Xue11,Hsu14}), the optical reference frame
  is probably shifted with respect to the radio calibrator reference
  frame used for ALMA astrometric calibration. We therefore corrected
  the optical positions in the GOOD-S region) by this offset.} Example
\textit{HST} and ALMA images of the X-ray sources are shown in
Fig.~\ref{Figure:Stamps} to demonstrate the quality of the optical and
ALMA data. The ALMA detection rate is comparable between X-ray sources
with photometric and spectroscopic redshifts, suggesting that
inaccurate redshifts are not a major reason for the
non-detections. Although our matching radii were $0.5^{\prime\prime}$
and $1^{\prime\prime}$, $\sim 80 \%$ of the ALMA counterparts lie
within $0.3^{\prime\prime}$ or less from the optical position of the
X-ray sources, including all of the 7 low-significance ALMA sources
giving us confidence that the majority are real sources.

\begin{figure}
	\includegraphics[width=\columnwidth]{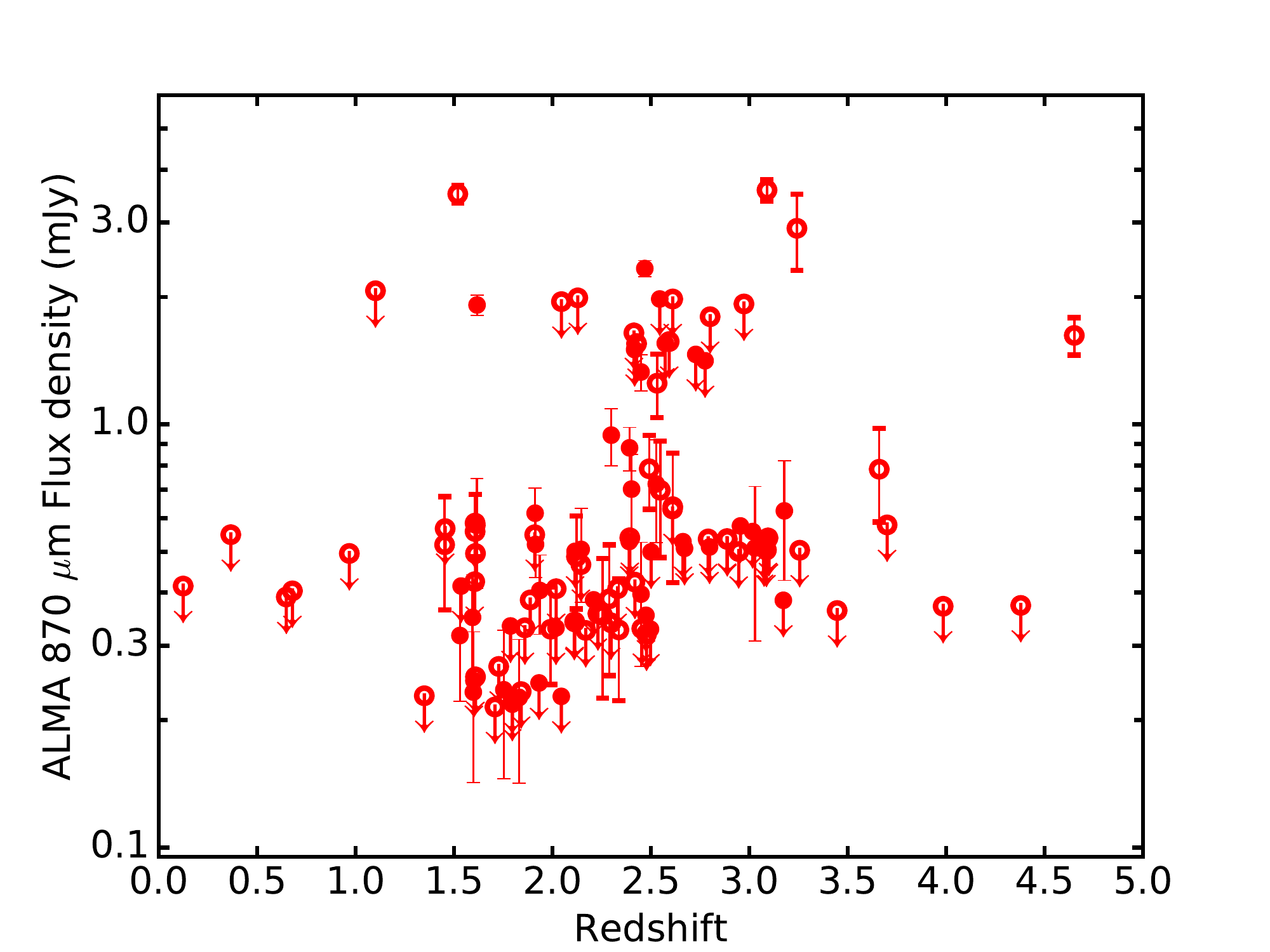}
   \caption{ALMA 870~$\mu$m flux density versus redshift for the X-ray
     detected that lie within our ALMA observations. The error bars
     represent the $1 \sigma$ error on the flux density.}
   \label{Figure:SED}
\end{figure}

The positions, redshifts and ALMA $870 \mu$m fluxes are summarised in
Tables \ref{Flux_table} \& \ref{Flux_tableb}. In addition to the 107
primary targets, there were a further 7 X-ray sources that
serendipitously lay within the field-of-view of the primary beam of
some of our ALMA maps. As a result we have ALMA coverage for 60 and 54
X-ray sources in the CDF-S and COSMOS fields respectively, covering a
$\lx$ range of $5 \times 10^{39} -10^{45} \ergss$ and a redshift range
of $z=$~0.1--4.6; see Fig.~ \ref{fig:Xlum} for the $z$--$\lx$
coverage. For the X-ray sources without an ALMA counterpart, we
calculated $3\sigma$ upper limits directly from the map. In Fig.~
\ref{Figure:SED} we show the ALMA 870~$\mu$m flux density versus
redshift for the 114 X-ray sources with ALMA coverage.

\begin{table*}
 \caption{X-ray selected sources observed with ALMA at $870
   \mu$m in CDF-S field. The columns show X-ray ID \citep[from][]{Hsu14},
   optical positions, ALMA positions, redshift (2 and 3 decimal places indicate
   photometric and spectroscopic redshifts, respectively), X-ray
   luminosity (rest-frame 2-10 keV), primary beam corrected ALMA
   fluxes, median baseline of the ALMA configuration, the RMS of
   the map containing the X-ray AGN and the observing ID.}  \resizebox{\textwidth}{!}{\begin{tabular}{@{}lcccccccccc@{}} 
\hline 
\hline 
X-ray ID&RA Optical&Dec Optical&RA ALMA&Dec ALMA&redshift& log$_{10}$&F$_{870 \mu m}$& Median baseline & RMS & Observing ID\\
 &(J2000)&(J2000)&(J2000)&(J2000)& & (L$_{2-10 \space \rm{keV}}$/erg s$^{-1}$)& (mJy) &  (m)&  (mJy) &\\
\hline 
$88$&$53.01019$&$-27.76674$&$53.01025$&$-27.76677$&$1.616$& $43.5$&$0.58\pm 0.17 $&220&$0.168$&2012.1.00869.S\\
$93$&$53.01265$&$-27.74724$& & &$2.573$& $43.5$&$<1.87$&393&$0.622$&2013.1.00884.S\\
$123$&$53.02794$&$-27.74866$& & &$2.33$& $42.7$&$<0.49$&220&$0.163$&2012.1.00869.S\\
$129$&$53.02961$&$-27.87481$& & &$3.45$& $43.8$&$<0.44$&91&$0.145$&2013.1.00884.S\\
$137$&$53.03333$&$-27.78258$& & &$2.610$& $43.9$&$<0.76$&220&$0.252$&2012.1.00869.S\\
$155$&$53.04094$&$-27.83607$& & &$2.02$&< $42.5$&$<0.49$&220&$0.163$&2012.1.00869.S\\
$156$&$53.04098$&$-27.83766$&$53.04108$&$-27.83774$&$4.65$& $43.6$&$1.62\pm 0.16 $&220&$0.163$&2012.1.00869.S\\
$158$&$53.04264$&$-27.86558$& & &$2.05$& $42.7$&$<2.34$&393&$0.780$&2013.1.00884.S\\
$163$&$53.04495$&$-27.77439$& & &$1.607$&< $42.3$&$<0.67$&220&$0.223$&2012.1.00869.S\\
$167$&$53.04567$&$-27.81557$& & &$1.46$& $43.1$&$<0.68$&220&$0.227$&2012.1.00869.S\\
$184$&$53.05220$&$-27.77477$& & &$1.605$& $42.3$&$<0.51$&220&$0.170$&2012.1.00869.S\\
$185$&$53.05233$&$-27.82728$&$53.05237$&$-27.82737$&$2.34$&< $42.4$&$0.33\pm 0.10 $&220&$0.104$&2012.1.00869.S\\
$195$&$53.05584$&$-27.81555$&$53.05584$&$-27.81566$&$1.45$& $42.9$&$0.52\pm 0.16 $&91&$0.155$&2013.1.00884.S\\
$199$&$53.05786$&$-27.83350$& & &$2.42$& $43.1$&$<1.80$&393&$0.601$&2013.1.00884.S\\
$211$&$53.06190$&$-27.85105$& & &$1.60$& $43.2$&$<0.30$&220&$0.099$&2012.1.00869.S\\
$215$&$53.06326$&$-27.69964$&$53.06326$&$-27.69971$&$2.402$& $43.1$&$0.70\pm 0.15 $&91&$0.146$&2013.1.00884.S\\
$221$&$53.06567$&$-27.87887$& & &$1.89$& $42.4$&$<0.46$&220&$0.154$&2012.1.00869.S\\
$230$&$53.06774$&$-27.92342$&$53.06781$&$-27.92361$&$3.98$& $43.7$&$0.43\pm 0.15 $&91&$0.149$&2013.1.00884.S\\
$249$&$53.07446$&$-27.84980$& & &$0.124$&< $39.8$&$<0.50$&220&$0.166$&2012.1.00869.S\\
$254$&$53.07600$&$-27.87816$& & &$2.801$& $43.1$&$<2.16$&393&$0.719$&2013.1.00884.S\\
$257$&$53.07640$&$-27.84866$& & &$1.536$& $43.7$&$<0.50$&220&$0.166$&2012.1.00869.S\\
$262$&$53.07846$&$-27.85986$&$53.07840$&$-27.86004$&$3.660$& $43.8$&$0.78\pm 0.20 $&220&$0.195$&2012.1.00869.S\\
$276$&$53.08270$&$-27.86657$&$53.08275$&$-27.86657$&$1.52$& $42.1$&$3.50\pm 0.16 $&220&$0.161$&2012.1.00869.S\\
$277$&$53.08313$&$-27.71198$& & &$2.21$& $43.4$&$<0.46$&91&$0.154$&2013.1.00884.S\\
$290$&$53.08732$&$-27.92955$& & &$2.55$& $43.6$&$<2.37$&393&$0.791$&2013.1.00884.S\\
$294$&$53.08918$&$-27.93047$& & &$2.611$& $43.3$&$<2.37$&393&$0.791$&2013.1.00884.S\\
$301$&$53.09229$&$-27.80316$&$53.09234$&$-27.80322$&$2.47$& $43.2$&$2.34\pm 0.10 $&220&$0.104$&2012.1.00869.S\\
$305$&$53.09379$&$-27.80131$& & &$2.42$& $42.7$&$<0.51$&220&$0.169$&2012.1.00869.S\\
$308$&$53.09392$&$-27.76772$& & &$1.727$& $43.6$&$<0.32$&220&$0.107$&2012.1.00869.S\\
$310$&$53.09403$&$-27.80413$&$53.09404$&$-27.80419$&$2.39$& $43.1$&$0.88\pm 0.10 $&220&$0.104$&2012.1.00869.S\\
$318$&$53.09636$&$-27.74506$&$53.09639$&$-27.74505$&$1.607$&< $42.2$&$0.58\pm 0.10 $&220&$0.099$&2012.1.00869.S\\
$320$&$53.09765$&$-27.71528$&$53.09771$&$-27.71537$&$2.145$& $42.8$&$0.56\pm 0.19 $&220&$0.186$&2012.1.00869.S\\
$326$&$53.10081$&$-27.71599$& & &$2.298$& $42.9$&$<0.41$&91&$0.136$&2013.1.00884.S\\
$344$&$53.10486$&$-27.70522$&$53.10487$&$-27.70532$&$1.617$& $43.4$&$1.92\pm 0.11 $&220&$0.105$&2012.1.00869.S\\
$351$&$53.10702$&$-27.71823$&$53.10709$&$-27.71834$&$2.532$& $44.1$&$1.25\pm 0.21 $&220&$0.214$&2012.1.00869.S\\
$359$&$53.10811$&$-27.75398$& & &$2.728$& $43.4$&$<1.76$&393&$0.585$&2013.1.00884.S\\
$371$&$53.11156$&$-27.76777$&$53.11157$&$-27.76782$&$3.24$& $43.5$&$2.91\pm 0.59 $&393&$0.594$&2013.1.00884.S\\
$386$&$53.11783$&$-27.73430$&$53.11797$&$-27.73438$&$3.256$&< $42.9$&$0.55\pm 0.20 $&220&$0.202$&2012.1.00869.S\\
$388$&$53.11858$&$-27.88480$& & &$2.13$& $42.7$&$<2.39$&393&$0.796$&2013.1.00884.S\\
$405$&$53.12283$&$-27.72280$& & &$1.609$& $42.7$&$<0.30$&220&$0.101$&2012.1.00869.S\\
$410$&$53.12409$&$-27.89120$&$53.12405$&$-27.89123$&$2.53$& $43.3$&$0.72\pm 0.20 $&220&$0.197$&2012.1.00869.S\\
$412$&$53.12436$&$-27.85163$& & &$3.700$& $44.1$&$<0.69$&220&$0.231$&2012.1.00869.S\\
$422$&$53.12557$&$-27.88646$&$53.12560$&$-27.88651$&$2.49$&< $42.7$&$0.79\pm 0.16 $&220&$0.156$&2012.1.00869.S\\
$423$&$53.12558$&$-27.88497$& & &$0.648$&< $41.4$&$<0.47$&220&$0.156$&2012.1.00869.S\\
$444$&$53.13403$&$-27.78096$& & &$2.39$& $43.4$&$<0.65$&220&$0.216$&2012.1.00869.S\\
$456$&$53.13799$&$-27.86825$& & &$3.17$& $43.1$&$<0.46$&91&$0.154$&2013.1.00884.S\\
$463$&$53.14102$&$-27.76673$& & &$1.910$&< $42.2$&$<0.66$&220&$0.219$&2012.1.00869.S\\
$466$&$53.14163$&$-27.81656$& & &$2.78$& $43.2$&$<1.70$&393&$0.566$&2013.1.00884.S\\
$470$&$53.14241$&$-27.76504$& & &$0.366$&< $40.7$&$<0.66$&220&$0.219$&2012.1.00869.S\\
$502$&$53.15118$&$-27.71608$& & &$0.968$& $41.9$&$<0.59$&220&$0.198$&2012.1.00869.S\\
$503$&$53.15119$&$-27.71373$& & &$1.609$&< $42.5$&$<0.59$&220&$0.198$&2012.1.00869.S\\
$509$&$53.15518$&$-27.74074$& & &$1.10$& $41.9$&$<2.48$&393&$0.828$&2013.1.00884.S\\
$522$&$53.15844$&$-27.77397$& & &$2.12$& $43.3$&$<0.60$&220&$0.200$&2012.1.00869.S\\
$528$&$53.16150$&$-27.85601$& & &$2.97$& $43.4$&$<2.31$&393&$0.770$&2013.1.00884.S\\
$534$&$53.16230$&$-27.71213$&$53.16240$&$-27.71222$&$4.379$& $43.5$&$0.44\pm 0.15 $&91&$0.149$&2013.1.00884.S\\
$535$&$53.16271$&$-27.74426$& & &$0.679$& $42.4$&$<0.48$&220&$0.162$&2012.1.00869.S\\
$574$&$53.17868$&$-27.80263$& & &$2.43$& $42.6$&$<1.86$&393&$0.621$&2013.1.00884.S\\
$593$&$53.18583$&$-27.80997$& & &$2.593$& $43.4$&$<1.88$&393&$0.628$&2013.1.00884.S\\
$633$&$53.20487$&$-27.91795$&$53.20489$&$-27.91800$&$2.30$& $43.4$&$0.94\pm 0.15 $&91&$0.146$&2013.1.00884.S\\
$677$&$53.24444$&$-27.90757$& & &$2.41$& $43.4$&$<1.97$&393&$0.658$&2013.1.00884.S\\
\hline 
\end{tabular}}
\label{Flux_table}
\end{table*}

\begin{table*}
 \caption{ X-ray selected sources observed with ALMA at $870
   \mu$m in COSMOS field. The columns show X-ray ID \citep[from][]{Marchesi16},
   optical positions, ALMA positions, redshift (2 and 3 decimal places indicate
   photometric and spectroscopic redshifts, respectively), X-ray
   luminosity (rest-frame 2-10 keV), primary beam corrected ALMA
   fluxes, median baseline of the ALMA configuration, the RMS of
   the map containing the X-ray AGN and the observing ID.}  \resizebox{\textwidth}{!}{\begin{tabular}{@{}lcccccccccc@{}} 
\hline 
\hline 
X-ray ID&RA Optical&Dec Optical&RA ALMA&Dec ALMA&redshift& log$_{10}$&F$_{870 \mu m}$& Median baseline & RMS & Observing ID\\
 &(J2000)&(J2000)&(J2000)&(J2000)& & (L$_{2-10 \space \rm{keV}}$/erg s$^{-1}$)& (mJy) &  (m)&  (mJy) &\\
\hline 
cid $434             $&$149.72072$&$2.34901$&$149.72067$&$2.34904$&$1.530$&$44.6$&$0.32\pm 0.10 $&91&$0.095$&2013.1.00884.S\\
cid $580             $&$149.85469$&$2.60694$& & &$2.11$&$44.5$&$<0.41$&91&$0.135$&2013.1.00884.S\\
cid $1620            $&$149.87585$&$2.69028$& & &$2.169$&$44.4$&$<0.39$&91&$0.130$&2013.1.00884.S\\
cid $558             $&$149.88252$&$2.50513$& & &$3.10$&$44.8$&$<0.64$&91&$0.214$&2013.1.00884.S\\
cid $330             $&$149.95583$&$2.02806$&$149.95575$&$2.02801$&$1.753$&$44.6$&$0.24\pm 0.09 $&91&$0.090$&2013.1.00884.S\\
cid $529             $&$149.98158$&$2.31501$& & &$3.017$&$44.6$&$<0.67$&91&$0.223$&2013.1.00884.S\\
cid $474             $&$149.99390$&$2.30146$& & &$1.796$&$44.5$&$<0.27$&91&$0.091$&2013.1.00884.S\\
cid $451             $&$150.00253$&$2.25863$&$150.00258$&$2.25864$&$2.450$&$44.6$&$0.40\pm 0.13 $&91&$0.129$&2013.1.00884.S\\
cid $1127            $&$150.01057$&$2.26939$& & &$2.390$&$44.1$&$<0.63$&91&$0.211$&2013.1.00884.S\\
cid $1205            $&$150.01070$&$2.33297$&$150.01079$&$2.33300$&$2.255$&$43.9$&$0.35\pm 0.13 $&91&$0.128$&2013.1.00884.S\\
cid $706             $&$150.01105$&$2.36766$& & &$2.11$&$43.9$&$<0.41$&91&$0.137$&2013.1.00884.S\\
cid $1246            $&$150.01559$&$2.44216$& & &$2.89$&$44.0$&$<0.64$&91&$0.214$&2013.1.00884.S\\
cid $532             $&$150.01985$&$2.34914$& & &$1.796$&$44.4$&$<0.26$&91&$0.087$&2013.1.00884.S\\
cid $1216            $&$150.02008$&$2.35365$& & &$2.663$&$44.1$&$<0.63$&91&$0.211$&2013.1.00884.S\\
cid $987             $&$150.02727$&$2.43472$& & &$1.860$&$44.0$&$<0.40$&91&$0.132$&2013.1.00884.S\\
cid $659             $&$150.03290$&$2.45859$& & &$2.045$&$44.0$&$<0.27$&91&$0.091$&2013.1.00884.S\\
cid $1214            $&$150.03677$&$2.35852$&$150.03680$&$2.35843$&$1.59$&$44.0$&$0.35\pm 0.09 $&91&$0.091$&2013.1.00884.S\\
cid $1143            $&$150.03682$&$2.25778$& & &$2.454$&$44.0$&$<0.39$&91&$0.132$&2013.1.00884.S\\
cid $351             $&$150.04262$&$2.06329$& & &$2.018$&$44.6$&$<0.40$&91&$0.132$&2013.1.00884.S\\
cid $708             $&$150.05225$&$2.36927$&$150.05226$&$2.36935$&$2.548$&$44.0$&$0.70\pm 0.21 $&91&$0.214$&2013.1.00884.S\\
cid $352             $&$150.05891$&$2.01518$& & &$2.498$&$44.6$&$<0.39$&91&$0.131$&2013.1.00884.S\\
cid $1247            $&$150.06346$&$2.42192$& & &$3.09$&$43.9$&$<0.61$&91&$0.202$&2013.1.00884.S\\
cid $1215            $&$150.06454$&$2.32905$&$150.06451$&$2.32912$&$2.450$&$44.1$&$1.33\pm 0.13 $&91&$0.132$&2013.1.00884.S\\
cid $459             $&$150.06467$&$2.19098$& & &$2.89$&$44.7$&$<0.64$&91&$0.215$&2013.1.00884.S\\
cid $960             $&$150.07462$&$2.30206$&$150.07455$&$2.30199$&$2.122$&$43.9$&$0.49\pm 0.12 $&91&$0.120$&2013.1.00884.S\\
cid $1219            $&$150.07600$&$2.26429$& & &$2.946$&$44.1$&$<0.60$&91&$0.200$&2013.1.00884.S\\
cid $72              $&$150.09154$&$2.39908$& & &$2.475$&$44.6$&$<0.42$&91&$0.141$&2013.1.00884.S\\
cid $85              $&$150.09653$&$2.29309$& & &$1.349$&$43.8$&$<0.27$&91&$0.091$&2013.1.00884.S\\
cid $467             $&$150.10201$&$2.10549$&$150.10194$&$2.10550$&$2.288$&$44.8$&$0.39\pm 0.13 $&91&$0.132$&2013.1.00884.S\\
cid $149             $&$150.10371$&$2.66577$& & &$2.955$&$44.7$&$<0.69$&91&$0.230$&2013.1.00884.S\\
cid $1144            $&$150.10477$&$2.24364$&$150.10469$&$2.24365$&$1.912$&$44.1$&$0.62\pm 0.09 $&91&$0.090$&2013.1.00884.S\\
cid $86              $&$150.11958$&$2.29591$&$150.11958$&$2.29595$&$1.831$&$44.3$&$0.23\pm 0.08 $&91&$0.084$&2013.1.00884.S\\
cid $87              $&$150.13304$&$2.30328$&$150.13309$&$2.30324$&$1.598$&$44.9$&$0.23\pm 0.09 $&91&$0.090$&2013.1.00884.S\\
cid $965             $&$150.15218$&$2.30785$&$150.15216$&$2.30779$&$3.178$&$44.2$&$0.62\pm 0.20 $&91&$0.197$&2013.1.00884.S\\
cid $914             $&$150.18001$&$2.23128$&$150.17992$&$2.23133$&$2.146$&$44.0$&$0.51\pm 0.13 $&91&$0.127$&2013.1.00884.S\\
cid $81              $&$150.18655$&$2.45533$&$150.18660$&$2.45530$&$1.991$&$44.0$&$0.33\pm 0.08 $&91&$0.085$&2013.1.00884.S\\
cid $121             $&$150.19180$&$2.54391$& & &$2.79$&$44.3$&$<0.64$&91&$0.214$&2013.1.00884.S\\
cid $917             $&$150.19263$&$2.21985$&$150.19260$&$2.21983$&$3.090$&$43.9$&$3.58\pm 0.20 $&91&$0.201$&2013.1.00884.S\\
cid $124             $&$150.20532$&$2.50293$& & &$3.07$&$44.3$&$<0.63$&91&$0.211$&2013.1.00884.S\\
cid $953             $&$150.21075$&$2.39147$& & &$3.095$&$44.1$&$<0.65$&91&$0.216$&2013.1.00884.S\\
cid $83              $&$150.21416$&$2.47502$& & &$3.075$&$44.5$&$<0.61$&91&$0.202$&2013.1.00884.S\\
cid $1085            $&$150.21634$&$1.98874$& & &$2.231$&$44.5$&$<0.43$&91&$0.143$&2013.1.00884.S\\
cid $915             $&$150.21909$&$2.27867$& & &$1.84$&$44.0$&$<0.28$&91&$0.093$&2013.1.00884.S\\
cid $976             $&$150.22527$&$2.35122$& & &$2.478$&$43.9$&$<0.38$&91&$0.128$&2013.1.00884.S\\
cid $954             $&$150.23180$&$2.36401$&$150.23178$&$2.36400$&$1.936$&$44.2$&$0.40\pm 0.09 $&91&$0.086$&2013.1.00884.S\\
cid $970             $&$150.23550$&$2.36176$& & &$2.501$&$44.6$&$<0.60$&91&$0.200$&2013.1.00884.S\\
cid $75              $&$150.24779$&$2.44215$&$150.24777$&$2.44216$&$3.029$&$44.7$&$0.51\pm 0.20 $&91&$0.203$&2013.1.00884.S\\
cid $31              $&$150.27214$&$2.23010$&$150.27217$&$2.23009$&$2.611$&$44.8$&$0.64\pm 0.22 $&91&$0.216$&2013.1.00884.S\\
cid $90              $&$150.28482$&$2.39505$& & &$1.932$&$44.4$&$<0.29$&91&$0.098$&2013.1.00884.S\\
cid $365             $&$150.28563$&$2.01459$& & &$2.671$&$44.5$&$<0.61$&91&$0.204$&2013.1.00884.S\\
cid $58              $&$150.32689$&$2.09415$& & &$2.798$&$44.5$&$<0.62$&91&$0.205$&2013.1.00884.S\\
cid $53              $&$150.34372$&$2.14067$& & &$1.787$&$44.2$&$<0.40$&91&$0.133$&2013.1.00884.S\\
cid $581             $&$150.35358$&$2.34220$& & &$1.708$&$44.5$&$<0.26$&91&$0.086$&2013.1.00884.S\\
cid $62              $&$150.37364$&$2.11203$&$150.37366$&$2.11203$&$1.914$&$44.5$&$0.52\pm 0.09 $&91&$0.086$&2013.1.00884.S\\
\hline 
\end{tabular}}
\label{Flux_tableb}
\end{table*}

%%%%%%%%%%%%%%%%%%%%%%%%%%%%%%%%%%%%%%%%%%%%%%%%%%

% Don't change these lines
\bsp	% typesetting comment
\label{lastpage}
\end{document}